\documentstyle[epsfig,prl,aps]{revtex}
\begin{document}
\draft \twocolumn[\hsize\textwidth\columnwidth\hsize\csname
@twocolumnfalse\endcsname

\title{Stability of a vacuum nonsingular black hole}
\author{Irina Dymnikova}
\address{Department of Mathematics and Computer Science,
University of Warmia and Mazury,\\
\.Zo{\l}nierska 14, 10-561 Olsztyn, Poland; e-mail:
irina@matman.uwm.edu.pl}
\author{Evgeny Galaktionov}
\address{A.F.Ioffe Physico-Technical Institute,
Politekhnicheskaja 26, St.Petersburg 194021, Russia}

\maketitle

\begin{abstract}
This is the first of series of papers in which we investigate stability of the
spherically symmetric space-time with de Sitter center.  Geometry, asymptotically
Schwarzschild for large $r$ and asymptotically de Sitter as $r\rightarrow 0$,
describes a vacuum nonsingular black hole for $m\geq m_{cr}$ and particle-like
self-gravitating structure for $m < m_{cr}$ where a critical value $m_{cr}$ depends on
the scale of the symmetry restoration to de Sitter group in the origin. In this paper
we address the question of stability of a vacuum non-singular black hole with de
Sitter center to external perturbations. We specify first two types of geometries with
and without changes of topology. Then we derive the general equations for an arbitrary
density profile and show that in the whole range of the mass parameter $m$ objects
described by geometries with de Sitter center remain stable under axial perturbations.
In the case of the polar perturbations we find criteria of stability and study in
detail the case of the density profile $\rho(r)=\rho_0 e^{-r^3/r_0^2 r_g}$ where
$\rho_0$ is the density of de Sitter vacuum at the center, $r_0=\sqrt{3/\kappa\rho_0}$
is de Sitter radius and $r_g$ is the Schwarzschild radius.
\end{abstract}


\vspace{0.2cm}

]

\section{Introduction}

The idea of replacing of a Schwarzschild singularity with de
Sitter vacuum goes back to 1965 papers of Sakharov \cite{sakharov}
who considered $p=-\rho$ as the equation of state for superhigh
density and of Gliner who interpreted $p=-\rho$ as corresponding
to a vacuum and suggested that it could be a final state in a
gravitational collapse \cite{gliner}.

In 1968 Bardeen presented the  spherically symmetric metric of the
same form as Schwarzschild and Reissner-Nordstr\"om metric,
describing a non-singular black hole (BH) without specifying the
behavior at the center \cite{bardeen}. The very important point
was noted in \cite{bardeen} for the first time: that the
considered space-time exhibits the smooth changes of topology.

Direct matching of Schwarzschild metric to de Sitter metric within
a short transitional space-like layer of the Planckian depth
\cite{markov,bernstein,farhi,shen,valera} results in metrics
typically with a jump at the junction surface.

The situation with transition to de Sitter as $r\rightarrow 0$,
has been analyzed in 1988 by Poisson and Israel who found
necessary to introduce a transitional layer of "non-inflationary
material" of uncertain depth at the characteristic scale $(r_0^2
r_g)^{1/3}$ ($r_0$ is de Sitter radius, and $r_g$ is the
Schwarzschild radius), where {\it geometry can be self-regulatory}
and describable semiclassically down a few Planckian radii by the
Einstein equations
with a source term representing vacuum polarization effects
\cite{werner}.

Generic properties of "noninflationary material" have been
considered in 1990 in Ref.\cite{me92}. For a smooth de
Sitter-Schwarzschild transition a source term satisfies
\cite{me92}
$$
T_t^t=T_r^r;~~~T_{\theta}^{\theta}=T_{\phi}^{\phi} \eqno(1.1)
$$
and the equation of state, following from $T^{\mu}_{\nu;\mu}=0$,
is
     $$p_r=-\rho;~~~p_{\perp}=-\rho-\frac{r}{2}\rho^{\prime}             \eqno(1.2)
     $$
Here $\kappa = 8\pi G$  (we adopted $c=1$ for simplicity),
$\rho(r)=T^t_t$ is the energy density, $p_r(r)=-T^r_r$ is the
radial pressure, and
$p_{\perp}(r)=-T_{\theta}^{\theta}=-T_{\phi}^{\phi}$ is the
tangential pressure for anisotropic perfect fluid \cite{tolman}.

The stress-energy tensor with the algebraic structure (1.1) has an
infinite set of comoving reference frames and is identified
therefore as describing a spherically symmetric vacuum
\cite{me92}, invariant under boosts in the radial direction and
defined by the symmetry of its stress-energy tensor (for review
\cite{spb,moscow,napoli,castle,observers}).

The exact analytical solution was found in 1990 for the case of
the density profile \cite{me92}
       $$\rho(r)=\rho_0 e^{-r^3/r_0^2 r_g};
       ~~r_0^2=3/\kappa\rho_0;~~r_g=2Gm                             \eqno(1.3)
       $$
which describes a vacuum asymptotically de Sitter as $r\rightarrow 0$  in a simple
semiclassical model for vacuum polarization in the spherically
symmetric gravitational field \cite{me96}.

In 1991 Morgan has considered a black hole in a simple model for
quantum gravity with quantum effects  represented by an upper
cutoff on the curvature, and obtained de Sitter-like past and
future cores replacing singularities \cite{morgan}. In 1992
Strominger demonstrated the possibility of natural, not {\it ad
hoc}, arising of de Sitter core inside a black hole in the model
of two-dimensional dilaton gravity conformally coupled to $N$
scalar fields \cite{strominger}.

In 1996 it was shown that in the course of Hawking evaporation a
vacuum nonsingular black hole evolves towards a self-gravitating
particle-like vacuum structure without horizons \cite{me96}, kind
of gravitational vacuum soliton called G-lump \cite{me2002}. The
form of temperature-mass diagram is generic for de
Sitter-Schwarzschild black hole \cite{me96} and dictated by the
Schwarzschild asymptotic and by the existence of two horizons -
when decreasing during evaporation mass reaches a certain critical
value $m_{cr}$ evaporation stops \cite{me96,me97}.

In 1997 in Ref.\cite{borde} it was shown that in a large class of
space-times that satisfy the Weak Energy Condition (WEC), the
existence of a regular black hole requires topology change.
Bardeen metric and the metric generated by the density profile
(1.3) belong to this class.

In 2000 studying the quantum gravitational effects by the
effective average action with the running Newton constant, and
improving Schwarzschild black hole with renormalization group,
Bonnano and Reuter \cite{BR2000} have constructed nonsingular
black hole metric and confirmed the results of \cite{me96}
concerning the form of temperature-mass diagram and the
fundamental fact that evaporation stops when the mass approaches
the critical value $m_{cr}$.

Also in 2000 the regular BH solution with a charged de Sitter core has been considered
by Kao \cite{Kao2000} with using the density profile (1.3) for distribution of a
charged material.

In the same year 2000 regular magnetic black hole and monopole
solutions are found  by Bronnikov \cite{kirmm} in Nonlinear
Electrodynamics (NED) coupled to gravity with the stress-energy tensor
of the structure (1.1).

Existence of regular electrically charged structures  in nonlinear
electrodynamics coupled to general relativity was proved recently
in Ref. \cite{me2004}, where it was shown that in NED coupled to
GR and satisfying WEC, regular charged structures must have de
Sitter center.

In 2001 the  non-singular quasi-black-hole model representing a
compact object without horizons, was constructed by Mazur and
Mottola \cite{MM2002} by extending the concept of Bose-Einstein
condensation to gravitational systems. An interior de Sitter
condensate phase is matched to an exterior Schwarzschild geometry
of arbitrary mass through a phase boundary of a small but finite
thickness with equation of state $p=\rho$.

In 2002 nonsingular  BH solution was found by Nashed
\cite{Nashed1} as a general solution of M\"oller tetrad theory of
gravitation by assuming the same specific form of the vacuum
stress-energy tensor as in Ref.\cite{me92} with the density
profile (1.3). Later it was extended to the case of teleparallel
theory of gravitation \cite{Nashed2}. Stability condition of
geodesic motion in the field of vacuum nonsingular black hole
described by the regular analytic solution \cite{me92} with the
density profile (1.3), has been considered in \cite{Nashed3}.

 Model-independent analysis of the Einstein spherically
symmetric minimally coupled equations has shown \cite{me2002,me2003} which geometry
they can describe  if certain general requirements are satisfied: (a) regularity of
density; (b) finiteness of the ADM mass; (c) Dominant Energy Condition (DEC) for
$T_{\mu\nu}$. These conditions lead to existence of regular structures with de Sitter
center including regular black holes without topological changes. The example of such
a case is the exact analytic solution \cite{me2004} describing in certain mass range a
regular charged black hole with de Sitter center.

The condition (c) can be loosed to (c2): weak energy condition for
$T_{\mu\nu}$ and regularity of pressures \cite{me2003}. WEC which
is contained in DEC, in both cases is needed for de Sitter
asymptotic at the center.

The requirements (a)-(c) either (a)-(c2) define the family of asymptotically flat
solutions with the regular center which includes the class of metrics asymptotically
de Sitter as $r\rightarrow 0$. A source term connects de Sitter vacuum in the origin
with the Minkowskli vacuum at infinity. Space-time symmetry changes smoothly from de
Sitter group at the center to the Poincare group at infinity through the radial boosts
in between, and the standard formula for the ADM mass relates it to both de Sitter
vacuum trapped inside an object and smooth breaking of space-time symmetry
\cite{me2002}.

Cases (c)-(c2) differ by behavior of the curvature scalar $R$. In the  case (c) it is
non-negative which evidences the existence of regular black holes without topological
changes. So, the class of metrics with de Sitter center includes two subclasses with
and without topological changes.

In this paper we specify conditions of existence of two types of
geometries with the de Sitter center and investigate stability of
configurations described by these geometries, by studying
perturbations in geometry via Einstein equations linearized about
the unperturbed space-time. Results are valid for geometries of
both types.

 This paper is organized as follows. In Sect.2 we
outline the conditions of existence and basic properties of
spherically symmetric geometries with de Sitter center. In Sect. 3
we introduce the basic equations describing axially symmetric
time-dependent perturbations of a spherically symmetric system
with de Sitter center. In Sect. 4 we prove stability of such a
system to axial perturbations. In Sect. 5 we analyze the case of
polar perturbations and derive criteria  of stability to these
perturbations. In Sect. 6 we apply the results to the case of the
vacuum nonsingular black hole with the density profile (1.3).
Section 7 contains summary and discussion.

\section{Spherically symmetric space-time with de Sitter center}

A static spherically symmetric line element can be written in the
form \cite{tolman}
     $$
     ds^2 = e^{\mu(r)}dt^2 - e^{\nu(r)} dr^2 - r^2 d\Omega^2
                                                                       \eqno(2.1)
     $$
where $d\Omega^2$ is the metric of a unit 2-sphere.
 Integration of the Einstein equations gives
    $$e^{-\nu(r)}=1-\frac{2GM(r)}{r};~~M(r)
    =4\pi\int_0^r{\rho(x)x^2dx}
                                                                              \eqno(2.2)
    $$
whose asymptotic for large $r$ is $e^{-\nu}=1-{2Gm}/{r}$, with the
mass parameter
   $$
   m=4\pi\int_0^{\infty}{\rho(r) r^2 dr}
                                                                               \eqno(2.3)
   $$
Requirement of regularity of density, $\rho_0=\rho(r\rightarrow 0)
< \infty$, leads to behavior of mass function $M(r)\sim{r^3}$ as
$r\rightarrow 0$ and thus $\nu(0)=0$.

To outline the  conditions of existence of spherically symmetric
space-time with de Sitter center, we need the Oppenheimer equation
\cite{oppi}
$$ T_t^t-T_r^r=p_r+\rho=
\frac{1}{\kappa}\frac{e^{-\nu}}{r}(\nu^{\prime}+\mu^{\prime})
                                                                   \eqno(2.4)
$$
The dominant energy condition $T^{00}\geq|T^{ab}|$ for each
$a,b=1,2,3$, holds if and only if \cite{HE}
      $\rho\geq0;~\rho + p_k\geq 0; ~  \rho-p_k\geq 0;~k=1,2,3$.
      It includes the weak energy condition which implies
      $\rho\geq0;~\rho + p_k\geq 0$.
      Together with the condition of regularity of density, DEC (via $p_k\leq\rho$)
      leads to $\mu^{\prime}+\nu^{\prime}=0$ as $r\rightarrow 0$ \cite{me2002}.

The same result can be achieved by requirement of regularity of pressure (the subclass
satisfying (a-c2)).

 In the limit
$r\rightarrow\infty$ the condition of finiteness of the mass (2.3) requires density
profile $\rho(r)$ to vanish at infinity quicker than $r^{-3}$. In the case (c) the
dominant energy condition requires pressures to vanish as $r\rightarrow\infty$. Then
$\mu^{\prime}=0$ and $\mu=$const at infinity. Rescaling the time coordinate allows one
to put the standard boundary condition $\mu\rightarrow 0$ as $r\rightarrow \infty$
which ensures asymptotic flatness needed to identify (2.3) as the ADM mass
\cite{wald}.

The same result can be achieved in the case (c2) by postulating
regularity of pressures including vanishing of $p_r$ at infinity
sufficient to get $\mu^{\prime}=0$ needed for asymptotic flatness.

The weak energy condition requires $\mu^{\prime}+\nu^{\prime}\geq
0$. The function $\mu+\nu$ is growing from $\mu=\mu(0)$ at $r=0$
to $\mu=0$ at $r\rightarrow\infty$, which gives $\mu(0)\leq 0$
\cite{me2002}.

The range of family parameter $\mu(0)$ includes $\mu(0)=0$. In
this case the function $\nu(r)+\mu(r)$ is zero at $r=0$ and at
$r\rightarrow\infty$, its derivative is non-negative (by WEC via
$\rho+p_k\geq 0$), it follows that  $\nu(r)=-\mu(r)$ everywhere.

A source term for this class of metrics corresponds to anisotropic
perfect fluid which satisfies the $r-$dependent equation of state
(1.2), and the weak energy condition $p_{\perp}+\rho\geq 0$
demands monotonic decreasing of a density profile,
$\rho^{\prime}\leq 0$ \cite{me2002}.

Behavior at $r\rightarrow 0$ is dictated by the WEC \cite{me2002}.
The equation of state near the center becomes $p=-\rho$ , which
gives de Sitter asymptotic as $r\rightarrow 0$
$$
 ds^2=\biggl(1-\frac{r^2}{r_0^2}\biggr)dt^2
     -\frac{dr^2}{\biggl(1-\frac{r^2}{r_0^2}\biggr)}-r^2d\Omega^2
                                                                         \eqno(2.5)
      $$
$$
T_{\mu\nu}=\rho_0 g_{\mu\nu};~~~~~~ r_0^2=\frac{3}{\Lambda}; ~~~
~~~ \Lambda=\kappa\rho_0
                                                                         \eqno(2.6)
$$
where $\rho_0=\rho(r\rightarrow 0)$ and $\Lambda$ is the
cosmological constant which appeared at the origin although was
not present in the basic equations.

Requirements (a-c) either (a-c2) lead thus to the existence of the
class of metrics
$$
ds^2=g(r)dt^2-\frac{dr^2}{g(r)}-r^2 d\Omega^2
                                                                       \eqno(2.7)
$$
$$
g(r)=1-\frac{R_g(r)}{r};  ~~~ R_g(r)=2G M(r);
                                                                        \eqno(2.8)
$$
$$
M(r) =4\pi\int_0^r{\rho(x)x^2dx}
                                                                       \eqno(2.9)
$$
which are  asymptotically de Sitter as $r\rightarrow 0$, and
asymptotically Schwarzschild at large $r$
$$
  ds^2=\biggl(1-\frac{r_g}{r}\biggr)-
     \frac{dr^2}{\biggl(1-\frac{r_g}{r}\biggr)}
     -r^2d\Omega^2; ~~~r_g=2Gm
                                                                    \eqno(2.10)
     $$
The weak energy condition defines the form of the metric function $g(r)$. In the
region $r>0$ it has only minimum and the geometry can have not more than two horizons:
a black hole horizon $r_{+}$ and an internal horizon $R_{-}$\cite{me2002}.

The scalar curvature $R$, proportional to the trace of stress-energy tensor $T$, is
proportional to $\rho-p_{\perp}$ for geometries satisfying (1.2), so that conditions
(a-c) and (a-c2) distinguish two types of geometries. In the case (a-c) satisfying DEC
requirement, scalar curvature remains non-negative, since DEC requires $\rho-p_k\geq
0$. The subclass satisfying (a-c) does not exhibit changes of topology by virtue of
DEC and can be specified as DEC-subclass. Dominant energy condition requires that each
principal pressure does not exceed the density which guarantees that speed of sound
can not exceed speed of light. In nonlinear electrodynamics coupled to gravity,
photons do not follow null geodesics of background geometry but propagate along null
geodesics of an effective geometry \cite{effective}, and propagation of photons
resembles propagation inside a dielectric medium \cite{mario}. In the case of the
regular NED structure satisfying DEC \cite{me2004}, it allows one to avoid problems
with speed of sound exceeding speed of light.

In the case (a-c2) scalar curvature $R(r)$ changes sign somewhere
and geometry experiences topological changes. This subclass
satisfying only  weak energy condition (needed in both cases for
de Sitter behavior at the center) can be specified as
WEC-subclass.

The case of the density profile (1.3) belongs to WEC-subclass
satisfying (c2). The metric function and the mass function are
given by \cite{me92}
$$
g(r)=1-\frac{r_g}{r}\biggl(1-e^{-r^3/r_0^2 r_g}\biggr); ~~
M(r)=m\biggl(1-e^{-r^3/r_0^2 r_g}\biggr)
                                                                   \eqno(2.11)
$$
Dominant Energy Condition is not satisfied so that a surface of
zero scalar curvature exists at which $R(r)=0$. Zero curvature
surface $r=r_s$ is shown in fig.1 together with two horizons (a
black hole event horizon $r_{+}$ and an internal horizon $r_{-}$),
and  the characteristic surface of any geometry with de Sitter
center: a zero-gravity surface $r=r_c$ beyond which the strong
energy condition of singularities theorems \cite{HE}, is violated
(zero-gravity surface is defined by
$2\rho+r\rho^{\prime}=0$\cite{me96}).
\begin{figure}
\vspace{-8.0mm}
\begin{center}
\epsfig{file=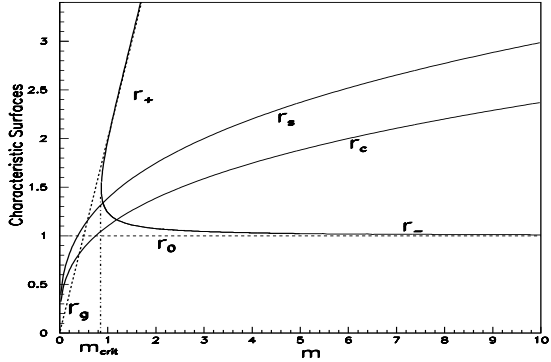,width=8.0cm,height=5.5cm}
\end{center}
\caption{Characteristic surfaces of a spherically symmetric space-time of WEC type
with de Sitter center.} \label{fig.1}
\end{figure}
Two horizons come together at the value of a mass parameter
$m_{cr}$, which puts a lower limit on a black hole mass (see
fig.2). For the case of a density profile (1.3) the critical mass
is given by \cite{me96}
$$
m_{cr}\simeq{0.3 m_{Pl}\sqrt{\rho_{Pl}/\rho_0}}
                                                                        \eqno(2.12)
$$

\begin{figure}
\vspace{-8.0mm}
\begin{center}
\epsfig{file=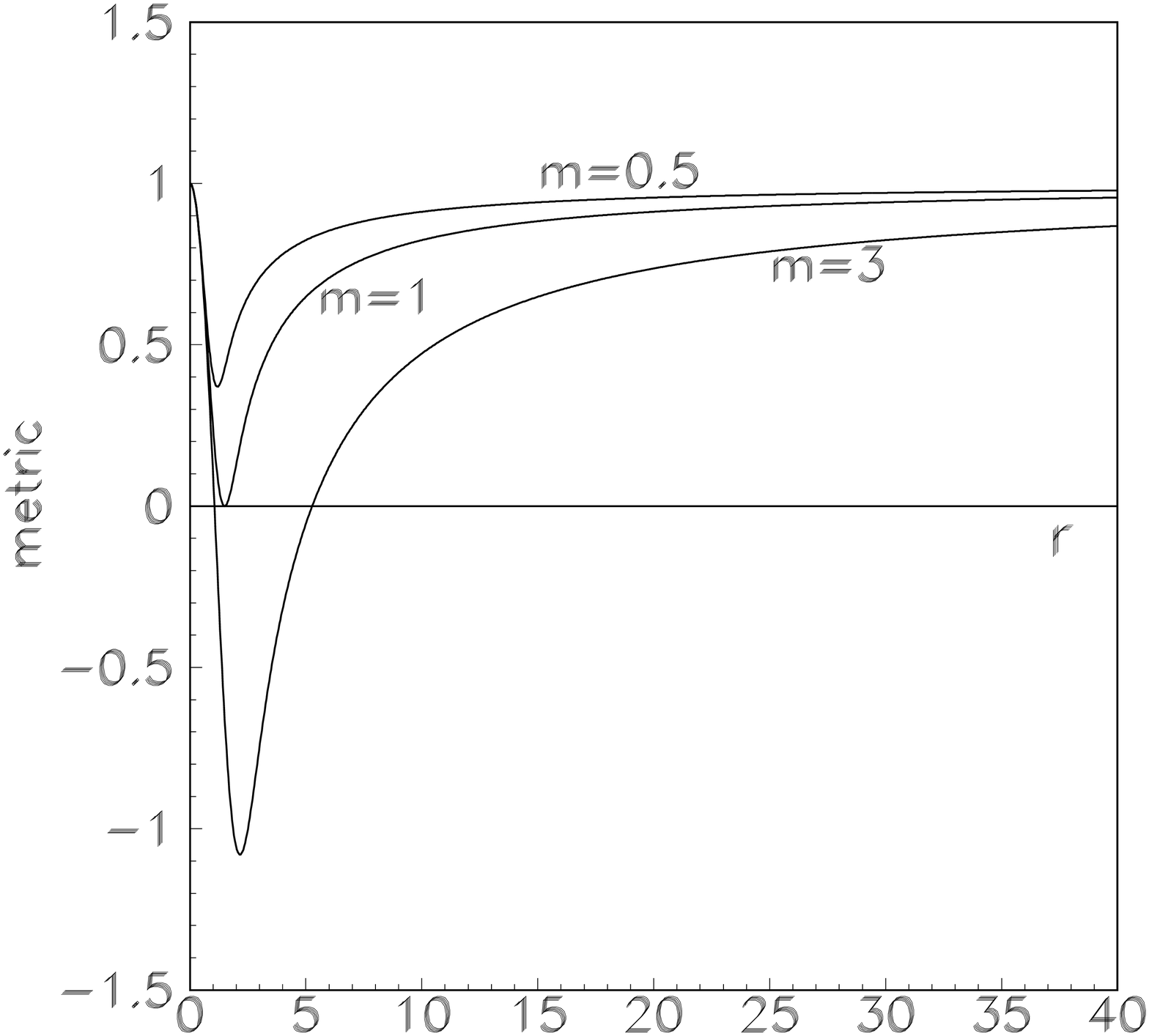,width=8.0cm,height=5.5cm}
\end{center}
\caption{ Metric function $g(r)$ for de Sitter-Schwarzschild configurations. Mass $m$
is normalized to $m_{cr}$. } \label{fig.2}
\end{figure}
Temperature-mass diagram is shown in fig.3. Its form does not
depend on particular choice of a density profile. Temperature
drops to zero at $m=m_{cr}$, while the Schwarzschild asymptotic
requires $T_{+}\rightarrow 0$ as $m\rightarrow\infty$. As a result
the temperature-mass diagram should have a maximum between
$m_{cr}$ and $m\rightarrow\infty$ \cite{me96}. In a maximum, at
$m=m_{cr2}$,  a specific heat is broken and changes sign
testifying to a second-order phase transition in the course of
Hawking evaporation \cite{me97}.

\begin{figure}
\vspace{-8.0mm}
\begin{center}
\epsfig{file=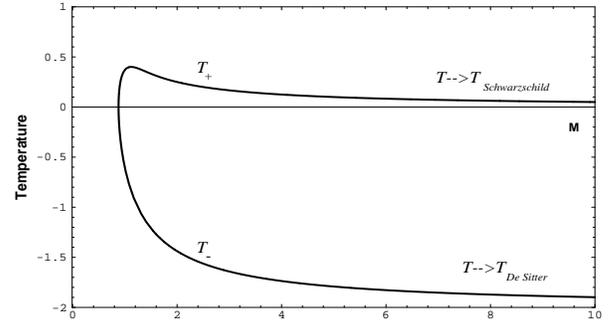,width=8.0cm,height=4.5cm,clip=}
\end{center}
\caption{ Temperature-mass diagram for a vacuum nonsingular black hole with de Sitter
center.} \label{fig.3}
\end{figure}

For $m\geq m_{cr}$ de Sitter-Schwarzschild geometry describes the vacuum nonsingular
black hole, and global structure of space-time shown in fig.4 \cite{me96}, contains an
infinite sequence of black and white holes whose future and past singularities are
replaced with regular cores ${\cal{RC}}$ asymptotically de Sitter as $r\rightarrow 0$
\cite{me96}.
\begin{figure}
\vspace{-8.0mm}
\begin{center}
\epsfig{file=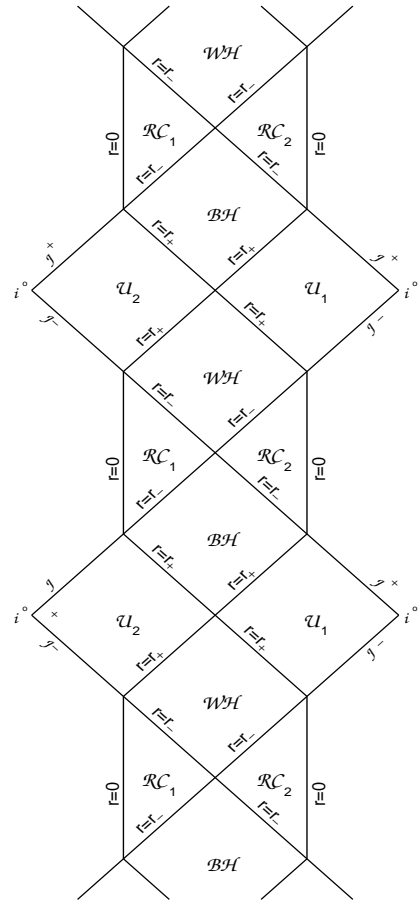,width=8.0cm,height=13.5cm}
\end{center}
\caption{Global structure of space-time of a vacuum nonsingular black hole with de
Sitter center \protect\cite{me96}.} \label{fig.4}
\end{figure}

\section{Basic equations for perturbations}

The perturbations of a spherically symmetric system are on essence
time-dependent axially symmetric modes; the reason is the absence
of a preferred axes in a spherically symmetric background
\cite{chandra}.

They are  described by the line element \cite{chandra}
$$
ds^2 = e^{2\nu}dt^2-e^{2\psi}(d\phi - \omega dt - q_2 dr -q_3
d\theta)^2
$$
$$
-e^{2\mu_{2}}(dr)^2 -e^{2\mu_{3}}(d\theta)^2,
                                                                    \eqno(3.1)
$$
in which metric functions $\nu, \psi, \mu_2, \mu_3, \omega, q_2,
q_3$ are functions only of $t, r, \theta$. They satisfy the
Einstein equations
$$
G_{ij}=R_{ij}-\frac{1}{2}g_{ij}R=-\kappa T_{ij},
                                                                    \eqno(3.2)
$$
where  the Ricci tensor is defined by $R_{ij}=g^{kl}R_{ikjl}$.

Non-zero components of stress-energy tensor read
$$
T_{tt}=e^{2\nu}\rho; ~~~  T_{\phi\phi}=e^{2\psi}p_{\phi};
$$
$$
T_{rr}=e^{2\mu_{2}}p_{r}; ~~ ~
T_{\theta\theta}=e^{2\mu_{3}}p_{\theta}.
                                                                    \eqno(3.3)
$$
where $p_{r}, p_{\theta}, p_{\phi}$ are the principal pressures.

We obtain the relevant perturbation equations by linearizing the
field equations around the spherically symmetric solution with de
Sitter center. This solution considered as a special case of the
line element (3.1) with
$$
\mu_{2}=-\nu(r);~~~ \psi=\ln(r\sin(\theta)); ~~~ \mu_{3}=\ln(r),
                                                                    \eqno(3.4)
$$
has the form (2.7) with
$$
g(r)=e^{2\nu(r)}=1+{C_{1}\over r}-{\kappa\over
r}\int{\rho(r)r^2dr},
                                                                    \eqno(3.5)
$$
The particular solution (3.5) is specified by the choice of the
constant $C_{1}$ which we choose in such a way to have unperturbed
metric given by (2.8)-(2.9).

\vskip0.1in

Our task is to investigate stability of the spherically symmetric
system with de Sitter center to external perturbations in general
case of a regular density profile $\rho(r)$.

The class of metrics with de Sitter center and a source term of
structure (1.1), is extended to the case of non-zero cosmological
constant ($\lambda < \Lambda)$ at infinity \cite{us97}
corresponding to extension of the Einstein cosmological term
$\Lambda g_{ik}$ to an $r-$dependent second rank symmetric tensor
$$
\Lambda_{ik}=\kappa T_{ik}
                                                                   \eqno(3.6)
$$
with the algebraic structure (1.1), connecting smoothly two de Sitter vacua with
different values of a cosmological constant \cite{me2000}. In this approach a constant
scalar $\Lambda$ associated with a vacuum density $\Lambda=\kappa\rho_{vac}$, becomes
a tensor component $\Lambda^t_t$ associated explicitly with a density component of a
perfect fluid tensor whose vacuum properties follow from its symmetry (1.1) and whose
variability follows from the Bianchi identities \cite{me2000,me2002}.

Here we investigate stability for the particular case when
$\lambda=0$ and spherically-symmetric space-time with de Sitter
center is asymptotically flat.

Since an anisotropic fluid with the stress-energy tensor of type
(1.1) admits identifying it as a vacuum-like medium
 associated with a time-evolving and spatially inhomogeneous
cosmological term \cite{me92,me2000,me2002,us2003}, we can write
the Einstein equations in the form
$$
G_{ik}+\Lambda_{ik}=0
                                                                 \eqno(3.7)
$$
(for discussion of where to put cosmological term see \cite{overduin,me2002}). Then
the quantities $\rho, p_k$ are treated as corresponding (in one-to-one way) components
of the variable cosmological term $\Lambda_t^t=\kappa\rho, \Lambda^k_k=-\kappa p_k$
\cite{me2000}.

Since we apply the approach of studying direct perturbations of
geometry via Einstein equations, we consider behavior of small
perturbations for both the metric tensor and a stress-energy
tensor associated with $\Lambda_{ik}$.

\vskip0.1in

 A general
perturbation of a background geometry will result in $\omega, q_2,
q_3$ becoming small quantities of the first order and the
functions $\nu, \mu_2, \mu_3, \psi$ and $\rho, p_k$ experiencing
small increments $\delta\nu, \delta\mu_2, \delta\mu_3, \delta\psi$
and $\delta\rho, \delta p_k$.

The perturbations leading to non-vanishing values of $\omega, q_2$
and $q_3$ induce a dragging of the inertial frame and impart a
rotation, for this reason they are called axial perturbations
\cite{chandra}.

Perturbations which do impart no rotation are called polar
perturbations \cite{chandra}. In the considered case they lead to
increments in $\nu, \mu_2, \mu_3, \psi$ and $\rho, p_k$.

The equations governing the axial and the polar perturbations
decouple.

\vskip0.1in

 Axial perturbations are governed by equations \cite{chandra}
$$
R_{r\phi}=R_{\theta\phi}=0
                                                             \eqno(3.8)
$$
The equations governing the polar perturbations read
$$
-R_{tr}=(\psi+\mu_{3})_{,rt}+ \psi_{,r}(\psi-\mu_{2})_{,t}
$$
$$
+ \mu_{3,r}(\mu_{3}-\mu_{2})_{,t}
 - \nu_{,r}(\psi+\mu_{3})_{,t}=0,
                                                              \eqno(3.9)
$$
$$
-R_{t\theta }=(\psi+\mu_{2})_{,\theta
t}+\psi_{,\theta}(\psi-\mu_{3})_{,t}
$$
$$
+\mu_{2,\theta}(\mu_{2}-\mu_{3})_{,t}
-\nu_{,\theta}(\psi+\mu_{2})_{,t}=0,
                                                              \eqno(3.10)
$$
$$
-R_{r\theta}=(\psi+\nu)_{,r
\theta}+\psi_{,r}(\psi-\mu_{2})_{,\theta}
$$
$$
+ \nu_{,r}(\nu-\mu_{2})_{,\theta}
-\mu_{3,r}(\psi+\nu)_{,\theta}=0,
                                                               \eqno(3.11)
$$
$$
G_{tt}=e^{-2\mu_{2}}[(\psi+\mu_{3})_{,rr}+ \psi_{,r}(\psi-\mu_{2}+\mu_{3})_{,r}
+\mu_{3,r}(\mu_{3}-\mu_{2})_{,r}]
$$
$$
+e^{-2\mu_{3}}[(\psi+\mu_{2})_{,\theta \theta}+ \psi_{,
\theta}(\psi+\mu_{2}-\mu_{3})_{, \theta}+ \mu_{2,\theta}(\mu_{2}-\mu_{3})_{,\theta}]
$$
$$
-e^{-2\nu}[\psi_{,t} (\mu_{2}+\mu_{3})_{,t}+ \mu_{2,t}\mu_{3,t}]=-\kappa e^{2\nu}\rho,
                                                                 \eqno(3.12)
$$
$$
G_{\phi\phi}=e^{-2\mu_{2}}[(\nu+\mu_{3})_{,rr}+\nu_{,r}(\nu-\mu_{2}+\mu_{3})_{,r}
+\mu_{3,r}(\mu_{3}-\mu_{2})_{,r}]
$$
$$
+e^{-2\mu_{3}}[(\nu+\mu_{2})_{,\theta \theta}+
\nu_{,\theta}(\nu+\mu_{2}-\mu_{3})_{,\theta}+\mu_{2,\theta}(\mu_{2}
-\mu_{3})_{,\theta}]
$$
$$
 -e^{-2\nu}[(\mu_{2}+\mu_{3})_{,tt}+
\mu_{2,t}(\mu_{2}+ \mu_{3}-\nu)_{,t}+\mu_{3,t}(\mu_{3}
-\nu)_{,t}]
$$
$$
=\kappa e^{2\psi} p_{\phi},                      \eqno(3.13)
$$
$$
G_{rr}=e^{-2\mu_{2}}[\psi_{,r}
(\nu+\mu_{3})_{,r}+\nu_{,r}\mu_{3,r}]
$$
$$
+e^{-2\mu_{3}}[(\psi+\nu)_{,\theta \theta}+\psi_{,\theta}(\psi
+\nu-\mu_{3})_{,\theta}+\nu_{,\theta}(\nu-\mu_{3})_{,\theta}]
$$
$$
-e^{-2\nu}[(\psi+\mu_{3})_{,tt}+\psi_{,t}(\psi+\mu_{3}-\nu)_{,t}+
\mu_{3,t}(\mu_{3}-\nu)_{,t}]
$$
$$
=\kappa e^{2\mu_2} p_{r},                              \eqno(3.14)
$$
$$
G_{\theta\theta}=e^{-2\mu_{2}}[(\psi+\nu)_{,rr}+\psi_{,r}(\psi+\nu-\mu_{2})_{,r}+
\nu_{,r}(\nu-\mu_{2})_{,r}]
$$
$$
+e^{-2\mu_{3}}[\psi_{,\theta}(\nu+\mu_{2})_{,\theta}+
\nu_{,\theta}\mu_{2,\theta}] -e^{-2\nu}[(\psi+\mu_{2})_{,tt}
$$
$$
+\psi_{,t}(\psi+\mu_{2}-\nu)_{,t} +\mu_{2,t}(\mu_{2}-\nu)_{,t}] =\kappa e^{2\mu_3}
p_{\theta}.                         \eqno(3.15)
$$
We perturb equations (3.8)-(3.15) up to the first order, and in
equations (3.12)-(3.15) we disturb both left and right hand sides.
As a result we obtain the linear system of 7 partial differential
equations for the polar perturbations, and the linear system of 2
equations for the axial perturbations.

\section{Axial Perturbations}

Axial perturbations corresponds to appearing of nonzero values
$\omega, q_2, q_3$ which vanish for unperturbed system.

They are governed by  the Einstein equations (3.8). This gives 2
equations for 3 functions which read
$$
\frac{e^{2\nu(r)}}{r^2 \sin^3{\theta}}[\sin^3{\theta}
(q_{2,\theta}-q_{3,r})]_{,\theta}=-(\omega_{,r}-q_{2,t})_{,t}
                                                                   \eqno(4.1)
$$
$$
\frac{e^{2\nu(r)}}{r^2}[r^2 e^{2\nu(r)}
(q_{2,\theta}-q_{3,r})]_{,r}=(\omega_{,\theta}-q_{3,t})_{,t}
                                                                   \eqno(4.2)
$$
Now we take
$$\omega(r,\theta,t)=\tilde{\omega}(r,\theta)e^{i\sigma t}$$
and similarly for $q_2,q_3$; in what follows we retain the same
symbols for the amplitudes of the perturbations which satisfy
equations
$$
\frac{e^{2\nu(r)}}{r^2 \sin^3{\theta}}\biggl[\sin^3{\theta}
(q_{2,\theta}-q_{3,r})\biggr]_{,\theta}=-(i\sigma
\omega_{,r}+\sigma^2 q_{2})
                                                                          \eqno(4.3)
$$
$$
\frac{e^{2\nu(r)}}{r^2}\biggl[r^2 e^{2\nu(r)}
(q_{2,\theta}-q_{3,r})\biggr]_{,r}=i\sigma \omega_{,\theta}+
\sigma^2 q_{3}
                                                                            \eqno(4.4)
$$
Expressing $\omega_{,r}$ from (4.3) and $\omega_{,\theta}$ from
(4.4), differentiating and equating
$\omega_{,r\theta}=\omega_{,\theta r}$ we get one equation
$$
r^4\frac{\partial}{\partial r}\biggl(\frac{e^{2\nu}}{r^2}\frac{\partial Q}{\partial
r}\biggr) + \sin^3{\theta}\frac{\partial}{\partial\theta}
\biggl(\frac{1}{\sin^3{\theta}}\frac{\partial Q}{\partial\theta}\biggr)+\sigma^2 r^2
e^{-2\nu} Q=0
                                                                           \eqno(4.5)
$$
Here
$$
Q(r,\theta)=e^{2\nu} r^2 \sin^3{\theta} (q_{2,\theta}-q_{3,r})
                                                                          \eqno(4.6)
$$
For the Schwarzschild metric, eq.(4.5) coincides with the
analogous Chandrasekhar equation (\cite{chandra}, Ch.4, eq.(18)).
Separating variables by $Q(r,\theta)=R(r)\Theta(\theta)$, we get
$$
r^2 e^{2\nu}\frac{d}{dr}\biggl(\frac{e^{2\nu}}{r^2}\frac{dR}{dr}\biggr) -\lambda
\frac{e^{2\nu}}{r^2}R + \sigma^2 R=0
                                                                          \eqno(4.7)
$$
$$\frac{d}{d\theta}\biggl(\frac{1}
{\sin^3{\theta}}\frac{d\Theta}{d\theta}\biggr)
+\frac{\lambda}{\sin^3{\theta}} \Theta=0
                                                                          \eqno(4.8)
$$
Solutions to (4.8) are Gegenbauer polynomials
$$
Q_l(\theta)=C^{-\frac{3}{2}}_{l+2}(\theta)= (P_{l,\theta\theta}-P_{l,\theta}
ctg\theta)\sin^2{\theta}
                                                                            \eqno(4.9)
$$
which gives
$$
\lambda_l=(l+2)(l-1);~~l=2,3,...
                                                                           \eqno(4.10)
 $$
General solution can be written in the form
$$
Q(r,\theta)=\sum_{l=2}^{\infty}{R_l(r)\Theta_l(\theta)}
                                                                             \eqno(4.11)
$$
Equation for $R_l(r)$ reads
$$
r^2 e^{2\nu}\frac{d}{dr}\biggl(\frac{e^{2\nu}}
{r^2}\frac{dR_l}{dr}\biggr) -\frac{e^{2\nu}} {r^2}(l+2)(l-1)R_l +
\sigma^2_l R_l=0
                                                                              \eqno(4.12)
$$
In "tortoise" coordinate $r_*=\int{dr/g(r)}$, we get Schr\"odinger
equation for $Z_l(r_*)=r^{-1}R_l(r_{*})$
$$
\biggl(\frac{d^2}{d r_*^2} +\sigma^2_l\biggr)Z_l=V_lZ_l
                                                                               \eqno(4.13)
$$
where the potential is given by
$$
V_l(r)=\frac{e^{2\nu}}{r^2}(\mu^2+2 e^{2\nu}-2 r \nu_{,r} e^{2\nu})
                                                                               \eqno(4.14)
$$
with
$$
\mu^2=(l+2)(l-1)
 \eqno(4.15)
$$

For the Schwarzschild geometry (4.14) coincides with the
 Regge-Wheeler potential (\cite{chandra}, Ch.4, eq.(28)).

With the asymptotic behavior of (4.14)
$$
V_{l} \rightarrow \frac{l(l+1)}{r^2} \quad{as~~~} r\rightarrow
r_{*}\rightarrow \infty
$$
$$
V_{l} \rightarrow const~~  e^{g'(r_{+})r_{*}} \quad{as~~~}
r_{*}\rightarrow -\infty
                                                         \eqno(4.16)
$$
solutions of (4.13) have asymptotic $e^{\pm i\sigma r_{*}}$ as $r_{*}\rightarrow \pm
\infty$ as in the case of the Schwarzschild geometry \cite{chandra}. For real $\sigma$
they describe propagation of ingoing and outgoing waves through one-dimensional
potential barrier, so that we have to look for solutions to (4.13), which satisfy the
boundary conditions \cite{chandra}
$$
Z_l\rightarrow e^{+i\sigma_l r_{*}} +R_l(\sigma_l) e^{-i\sigma_l
r_{*}}~~~(r_{*} \rightarrow +\infty)
$$
$$
Z_l\rightarrow T_l(\sigma_l) e^{+i\sigma_l r_{*}} ~~~~
(r_{*}\rightarrow -\infty)
                                                                           \eqno(4.17)
$$
These boundary conditions tell us that each $l$ component of
an incident wave of the unit amplitude coming from infinity gives
rise to a reflected wave of amplitude $R_l(\sigma_l)$ at infinity
and a transmitted wave of amplitude $T_l(\sigma_l)$ at the horizon
\cite{chandra}.

If such solutions exist only for real values of the time parameter
 ${\sigma}_{l}$ and form complete basic set,  then any smooth initial
 perturbation defined at the finite interval of  $r_{*}$
 (with compact support), can be expanded on these functions,
  and since dependence of perturbations from
 time coordinate has the form $\exp(i{\sigma}_{l}t)$, this is followed by stability
 of geometry in question.

In terms of the metric function $g(r)$
$$
V_l(r)=\frac{g(r)}{r^2}(\mu^2+2 g- r g')
                                                                           \eqno(4.18)
$$
It is easily seen that $V_l(r)$ is constrained by the function
$$
V_l(r) \geq \frac{3 g}{r^2}\biggl(1+g+\frac{1}{3}\kappa\rho
r^2\biggr),
$$
which is certainly positive and presents the value of the
potential for the mode $l=2$.

\vskip0.1in

We see that axial perturbations are governed by one-dimensional
wave equation (4.13) with a non-negative  potential. In terms of a
one-dimensional Schr\"odinger equation, an unstable mode exists if
a potential has a bound state, which corresponds to the negative
eigenvalue $\sigma_l^2$.
 In the case of non-negative potential the system obeys the theorem \cite{ber}
 which guarantees the absence of negative eigenvalues in the standard
 one-dimensional Schr\"odinger equation with the non-negative potential.
 The absence of negative eigenvalues in the spectrum of (4.13)
guarantees the absence of exponentially growing modes.

For the case when the potential  is real, smooth and short-range,
standard theorems of quantum mechanics guarantee that
eigenfunctions of (4.13) form a complete set and any
square-integrable state function can be expanded on them
\cite{chandra}.

\vskip0.1in

We have however to be careful about the behavior of the solution to (4.13) in the
extremal regime near the double horizon  $r_{\pm}$ which satisfies $g(r_{\pm})=0;
 ~g'(r_{\pm})=0$.

To study the extreme case we introduce dimensionless variable $x$
by normalizing the variable $r$ to the characteristic scale
$(r_0^2 r_g)^{1/3}$.

Then the equation for the function $R_l$  reads
$$
x^2 g^2 R''_l + 2x\biggl(\frac{x g'}{2} -g\biggr) g R'_l +
({\sigma}^2_l x^2 -\mu^2 g)R_l=0
                                                                        \eqno(4.19)
$$
where ${\sigma}_l=\sigma_l (r_0^2 r_g)^{1/3}$. In what follows we
retain the notation $\sigma_l$ keeping in mind that it is
multiplied by $(r_0^2 r_g)^{1/3}$. (A multiplier is not essential
since in studying stability we are interested only in the sign of
$\sigma_l^2$.)

 Near the double horizon $x_{\pm}$, the metric function is
 $g(x)=g''(x_{\pm})(x-x_{\pm})^2/2+...$. Introducing the variable
$z=x-x_{\pm}$ and the notation $\gamma=g''(x_{\pm})/2$, we get the
limiting equation
$$ z^4R_{l,zz}+2z^3 R_{l,z}+\frac{\sigma_l^2}{\gamma^2} R_l=0
                                                                        \eqno(4.20)
$$
General solution to (4.20) (found by taking $R_l=w_l/z$),reads \cite{kamke}
$$
R_l(z)=C_{1l}cos\biggl(\frac{\kappa_l}{z}\biggr)
+C_{2l}sin\biggl(\frac{\kappa_l}{z}\biggr)
                                                                        \eqno(4.21)
$$
(with $\kappa_l^2={\sigma_l^2}/{\gamma^2}$), and tells us that axial perturbations are
restricted near double horizon.

In case without horizons asymptotic behavior of the potential at infinity and near
zero is $V_l={l(l+1)}/{r^2}$. This is behavior of centrifugal part of radial
Schr\"odinger operator in the spherically symmetric case.

\vskip0.1in

 We see that potential for axial perturbations is smooth,
short-range and positive for  all types of configurations
described by spherically symmetric geometry with de Sitter center.
We conclude  that geometry with de Sitter center is stable to
axial perturbations.

\section{Polar perturbations}

\subsection{General equations}

Linearizing the system (3.9)-(3.15) about the background geometry, we get the system
describing polar perturbations
$$
\biggl[(\delta \psi+\delta \mu_{3})_{,r}- \biggl(\nu_{,r}-{1\over
r}\biggr)(\delta \psi +\delta \mu_{3})- {2\over r}\delta
\mu_{2}\biggr]_{,t}=0,
                                                                      \eqno(5.1)
$$
$$
[(\delta \psi+\delta \mu_{2})_{,\theta}
+ ctg(\theta) (\delta \psi
- \delta \mu_{3})]_{,t}=0,
                                                   \eqno(5.2)
$$
$$
[\delta \psi_{,\theta}+ctg(\theta)(\delta \psi-\delta
\mu_{3})]_{,r} +\delta \nu_{,r \theta}
$$
$$+ \biggl(\nu_{,r}-{1\over
r}\biggr)\delta \nu_{,\theta} -\biggl(\nu_{,r}+{1\over
r}\biggr)\delta \mu_{2,\theta}=0,
                                                        \eqno(5.3)
$$
$$
-e^{2\nu}\biggl[(\delta \psi+\delta \mu_{3})_{,rr}
+\biggl(\nu_{,r}+{3\over r}\biggr)(\delta \psi+\delta
\mu_{3})_{,r}
$$
$$
- {2\over r}\delta \mu_{2,r}-2\biggl({2\over r}\nu_{,r}+ {1\over
r^2}\biggr)\delta \mu_{2}\biggr] -{1\over r^2}[\delta
\psi_{,\theta \theta}+ ctg(\theta)(2\delta \psi-\delta
\mu_{3})_{,\theta}
$$
$$
+2\delta \mu_{3}+\delta \mu_{2,\theta \theta}
+ ctg(\theta)\delta
\mu_{2,\theta}]=\kappa \delta \rho,
                                                        \eqno(5.4)
$$
$$
e^{2\nu}\biggl[(\delta \nu+\delta \mu_{3})_{,rr}
+\biggl(3\nu_{,r}+{1\over r}\biggr)\delta \nu_{,r}+
2\biggl(\nu_{,r}+{1\over r}\biggr)\delta \mu_{3,r}
$$
$$
-\biggl(\nu_{,r}+{1\over r}\biggr)\delta
\mu_{2,r}-2\biggl(\nu_{,rr}+ 2\nu_{,r}^2+{2\over
r}\nu_{,r}\biggr)\delta \mu_{2}\biggr]
$$
$$
 +{1\over r^2}(\delta \nu+\delta \mu_{2})_{,\theta \theta}-
e^{-2\nu}(\delta \mu_{2} +\delta \mu_{3})_{,tt}=\kappa\delta
p_{\phi},
                                                         \eqno(5.5)
$$
$$
e^{2\nu}\biggl[{2\over r}\delta \nu_{,r}+\biggl(\nu_{,r}+{1\over
r}\biggr) (\delta \psi+\delta \mu_{3})_{,r}- 2\biggl({2\over
r}\nu_{,r}+{1\over r^2}\biggr)\delta \mu_{2}\biggr]
$$
$$
 +{1\over r^2}[\delta \psi_{,\theta \theta}+
ctg(\theta)(2\delta \psi-\delta \mu_{3})_{,\theta} +2\delta
\mu_{3}+\delta \nu_{,\theta \theta}+ ctg(\theta)\delta
\nu_{,\theta}]
$$
$$
 -e^{-2\nu}(\delta \psi+\delta \mu_{3})_{,tt}=
\kappa \delta p_{r},
                                                             \eqno(5.6)
$$
$$
e^{2\nu}\biggl[(\delta \nu+\delta \psi)_{,rr}
+\biggl(3\nu_{,r}+{1\over r}\biggr)\delta \nu_{,r}+
2\biggl(\nu_{,r}+{1\over r}\biggr)\delta \psi_{,r}
$$
$$
-\biggl(\nu_{,r}+{1\over r}\biggr)\delta
\mu_{2,r}-2\biggl(\nu_{,rr}+ 2\nu_{,r}^2+{2\over
r}\nu_{,r}\biggr)\delta \mu_{2}\biggr]
$$
$$
 +{1\over
r^2}ctg(\theta)(\delta \nu+\delta \mu_{2})_{,\theta}
-e^{-2\nu}(\delta \mu_{2}+\delta \psi)_{,tt}=\kappa \delta
p_{\theta}.
                                                                  \eqno(5.7)
$$

The system (5.1)-(5.7) is the system of 7 linear partial differential equations of the
first order for 8 quantities:
 4 small perturbations of metric tensor and 4 small perturbations
of stress-energy tensor (which is in considered case can be associated with a variable
cosmological term) whose unperturbed components are related by the equation of state,
in our case (1.2). To investigate this system we should make an assumption concerning
perturbation of $p_r$  valid for the case of small perturbations. Since for the
background geometry  we have $p_{r}=-\rho$, i.e. $p_r=p_r(\rho)$, we can assume (see,
e.g., \cite{LL})
$$
\delta p_{r}={dp_{r}\over d\rho} \delta \rho.
                                                      \eqno(5.8)
$$
which results in
$$
\delta p_{r}= - \delta \rho.
                                                     \eqno(5.9)
$$
 The possibility to
connect perturbations $\delta p_r$ and $\delta \rho$ is implied by
our system which contains 7 equations for 8 functions. The
relation (5.9) is valid only for small perturbations, since only
in this case the relation (5.8) is valid. So, if we prove that the
system is stable, i.e. growing perturbation modes are absent, this
will justify the validity of (5.9).

 Taking into account  (5.4) and (5.6), the equation  (5.9)
can be written in the form
$$
e^{2\nu}\biggl[-(\delta \psi+\delta \mu_{3})_{,rr} -{2\over
r}(\delta \psi+\delta \mu_{3})_{,r}+ {2\over r}\delta
\nu_{,r}+{2\over r}\delta \mu_{2,r}\biggr]
$$
$$
+{1\over r^2}\biggl[\delta \nu_{,\theta \theta}+ ctg(\theta)
\delta \nu_{,\theta} -\delta \mu_{2,\theta \theta}-
ctg(\theta)\delta \mu_{2,\theta}\biggr]
$$
$$
-e^{-2\nu}(\delta \psi+\delta\mu_{3})_{,tt}=0.
                                                              \eqno(5.10)
$$
In this way we obtain the system of 7 equations for 7 unknown functions which splits
into uniform system of 4 linear partial differential equations (5.1), (5.2), (5.3),
(5.10) for 4 small perturbations  of the metric tensor, $\delta \nu(r,\theta,t),\delta
\mu_{2}(r,\theta,t), \delta \mu_{3}(r,\theta,t),\delta \psi(r,\theta,t)$; and 3 linear
algebraic equations (5.6), (5.5), (5.7), determining expressions for $\delta
p_{r},\delta p_{\phi},\delta p_{\theta}$ through expressions for metric perturbations.

\vskip0.1in

The problem ultimately reduces to investigation of the uniform
linear system (5.1), (5.2), (5.3), (5.10).

Following Chandrasekhar \cite{chandra} we assume the time dependence $e^{i\sigma t}$
which  corresponds to the Fourier analysis of perturbations. The variables $r$ and
$\theta$ are separated by the Friedman substitutions \cite{chandra}.

As a result we present perturbations as series
$$
\delta \nu(r,\theta,t)=
\sum_{l=2}^{+\infty}N_{l}(r)P_{l}(cos\theta) e^{i\sigma_{l}t},
                                                                     \eqno(5.11)
$$
$$
\delta \mu_{2}(r,\theta,t)=
\sum_{l=2}^{+\infty}L_{l}(r)P_{l}(cos\theta) e^{i\sigma_{l}t},
                                                                      \eqno(5.12)
$$
$$
\delta \mu_{3}(r,\theta,t)=\sum_{l=2}^{+\infty}[T_{l}(r)P_{l}(cos\theta)
$$
$$
+ V_{l}(r)P_{l,\theta \theta}(cos\theta)] e^{i\sigma_{l}t},
                                                                       \eqno(5.13)
$$
$$
\delta \psi(r,\theta,t)=\sum_{l=2}^{+\infty}[T_{l}(r)P_{l}(cos\theta)
$$
$$
+ V_{l}(r)P_{l,\theta}(cos\theta)ctg\theta] e^{i\sigma_{l}t},
                                                                        \eqno(5.14)
$$
$$
\delta
\rho(r,\theta,t)=\sum_{l=2}^{+\infty}C_{l}(r)P_{l}(cos\theta)
e^{i\sigma_{l}t},
                                                                          \eqno(5.15)
$$
$$
\delta p_{r}(r,\theta,t)
=\sum_{l=2}^{+\infty}D_{l}(r)P_{l}(cos\theta) e^{i\sigma_{l}t},
                                                                             \eqno(5.16)
$$
$$
\delta p_{\phi}(r,\theta,t)
=\sum_{l=2}^{+\infty}[E_{l}(r)P_{l}(cos\theta)
$$
$$
+H_{l}(r)P_{l,\theta \theta}(cos\theta)] e^{i\sigma_{l}t},
                                                                             \eqno(5.17)
$$
$$
\delta p_{\theta}(r,\theta,t)
=\sum_{l=2}^{+\infty}[E_{l}(r)P_{l}(cos\theta)
$$
$$
+ H_{l}(r)P_{l,\theta}(cos\theta)ctg\theta] e^{i\sigma_{l}t}.
                                                                           \eqno(5.18)
$$

Let us introduce  the function $X_l$ which will be useful in our
further reductions
$$
X_{l}(r)= n V_{l}(r),
                                                                 \eqno(5.19)
$$
where
$$
n={l(l+1)\over 2} -1; ~~  l=2,3,...;  ~~ n=2,5,9,....
$$
Using the properties of the Legendre polynomials
$$
(sin{\theta} P_{l,\theta})_{,\theta} +l(l+1)sin{\theta}
P_{l}(cos{\theta})=0,
$$
$$
P_{l,\theta \theta}+P_{l,\theta}ctg{\theta}
=-l(l+1)P_{l}(cos{\theta}),
$$
we get from Eqs (5.1)-(5.3), (5.10) after some algebra, the
following relations between amplitudes
$$
T_{l}(r)=V_{l}(r)-L_{l}(r);
$$
$$
(X_{l}(r)+L_l(r))_{, r} -\biggl(\nu_{, r}-{1\over r}\biggr)(X_l(r)
+L_l(r)) +\frac{1}{r} L_l(r)=0;
$$
$$
(T_{l}(r)-V_{l}(r))_{,r}-\biggl(\nu_{, r}+{1\over
r}\biggr)L_{l}(r) + N_{l,r}(r)+\biggl(\nu_{,r}-{1\over
r}\biggr)N_{l}(r)=0;
$$
$$
e^{2\nu}\biggr[\biggl(X_l(r)+L_l(r)\biggr)_{,r,r}
+\frac{1}{r}\biggl(N_l(r)+2X_l(r)+3L_l(r)\biggr)_{,r}\biggr]
$$
$$
-\frac{l(l+1)}{2r^2}\biggl(N_l(r)-L_l(r)\biggr)
-e^{-2\nu}\sigma_l^2\biggl(X_l(r)+L_l(r)\biggr)=0
$$
With using these relations we transform ultimately our starting
system to the system of 3 differential equations in the normal
form for the functions $N_l(r), L_l(r), X_l(r)$
$$
N_{l,r}=(n+1)a1N_{l}+(\nu_{,r}+b1
$$
$$
-(n+1)a1+\sigma_{l}^2c1)L_{l} +(b1+\sigma_{l}^2c1)X_{l},
                                                                  \eqno(5.20a)
$$
$$
L_{l,r}=\biggl(\nu_{,r}-{1\over r}+(n+1)a1\biggr)N_{l}+\biggl(-{1\over r}+ b1
$$
$$
-(n+1)a1+\sigma_{l}^2c1\biggr)L_{l}+(b1+\sigma_{l}^2c1))X_{l},
                                                                    \eqno(5.20b)
$$
$$
X_{l,r}=\biggl(-\nu_{,r}+{1\over r}-(n+1)a1\biggr)N_{l}
$$
$$
+\biggl(\nu_{,r}-{1\over r} -b1+(n+1)a1-\sigma_{l}^2c1\biggr)L_{l}
$$
$$
+\biggl(\nu_{,r}-{1\over r}-b1-\sigma_{l}^2c1\biggr)X_{l},
                                                                    \eqno(5.20c)
$$
where
$$
a1(r)=\frac{e^{-2\nu(r)}}{r}; ~
b1(r)=-r\nu_{,rr}-r\nu_{,r}^2;~
c1(r)=r e^{-4\nu(r)},
                                        \eqno(5.21)
$$
and 4 equations which define amplitudes $D_{l}(r)$, $E_{l}(r)$,
$H_{l}(r)$, and $C_{l}(r)$ through solutions of (5.20)
$$
 D_{l}(r)={2\over \kappa}\biggl[e^{2\nu}\biggl[{1\over r}N_{l,r}-
\biggl(\nu_{,r}+{1\over r}\biggr)(X_{l}+L_{l})_{,r}
$$
$$
-\biggl({2\over r}\nu_{,r}+ {1\over r^2}\biggr)L_{l}\biggr]
-{1\over r^2}[X_{l}- n L_{l}+(n+1)N_{l}]
$$
$$
-\sigma_{l}^2 e^{-2\nu}(X_{l}+L_{l})\biggr],
                                                                        \eqno(5.22)
$$
$$
E_{l}(r)={1\over
\kappa}\biggl[e^{2\nu}\biggl[\biggl(N_{l}-L_{l}+{1\over n}
X_{l}\biggr)_{,rr}
$$
$$
+\biggl(3\nu_{,r}+{1\over r}\biggr)N_{l,r}+
2\biggl(\nu_{,r}+{1\over r}\biggr)\biggl({1\over n}
X_{l}-L_{l}\biggr)_{,r}
$$
$$
-\biggl(\nu_{,r}+{1\over r}\biggr)L_{l,r}- 2\biggl(\nu_{,rr}+
2\nu_{,r}^2+{2\over r}\nu_{,r}\biggr)L_{l}\biggr]
$$
$$
+ \sigma_{l}^2 e^{-2 \nu} {1\over n} X_{l}\biggr],
                                                                         \eqno(5.23)
$$
$$
H_{l}(r)={1\over {n \kappa}}\biggl[e^{2\nu}\biggl[ X_{l,rr}+
2\biggl(\nu_{,r}+{1\over r}\biggr)X_{l,r}\biggr]
$$
$$
+{n\over r^2}(N_{l}+L_{l})+ \sigma_{l}^2 e^{-2\nu} X_{l}\biggr],
                                                                           \eqno(5.24)
$$
$$
C_{l}(r)=-D_{l}(r),
                                                                            \eqno(5.25)
$$
Let us now introduce the dimensionless variables
$$ x={r\over r_{1}};~~\rho\rightarrow \frac{\rho}{\rho_0}; ~
\quad{where}~  ~r_{1}^3=r_{0}^2 r_{g}
                                                                 \eqno(5.26)
$$
and the characteristic parameter
$$
\alpha={r_{g}\over r_{1}}
                                                                           \eqno(5.27)
  $$
In these notations the unperturbed solution  (2.7) reads
$$
ds^2=g(x)dt^2-\frac{dx^2}{g(x)}-x^2d\Omega^2
$$
$$
g(x)=1-\frac{\alpha M(x)}{x}; ~~ ~M(x)=3\int_0^x{\rho(q)q^2dq}
                                                                         \eqno(5.28)
$$
In terms of $g(x)$  our basic system (5.20) takes the form
$$
N_{l,x}={(n+1)\over xg} N_{l}+\biggl[{1\over 2}{g'\over g}
+{x\over 4}\biggl({g'\over g}\biggr)^2-{x\over 2}{g''\over g}
$$
$$
 -{(n+1)\over xg}+
\sigma_{l}^2 {x\over g^2}\biggr] L_{l}+ \biggl[{x\over
4}\biggl({g'\over g}\biggr)^2-{x\over 2}{g''\over g} +
\sigma_{l}^2 {x\over g^2}\biggr] X_{l}
                                                               \eqno(5.29a)
$$
$$
L_{l,x}=\biggl[{1\over 2}{g'\over g}-{1\over x}+{(n+1)\over
xg}\biggr] N_{l} +\biggl[-{1\over x}+{x\over 4}\biggl({g'\over
g}\biggr)^2-{x\over 2} {g''\over g}
$$
$$
-{(n+1)\over xg}+ \sigma_{l}^2 {x\over g^2}\biggr] L_{l}+
\biggl[{x\over 4}\biggl({g'\over g}\biggr)^2-{x\over 2}{g''\over
g} +\sigma_{l}^2 {x\over g^2}\biggr] X_{l},
                                                             \eqno(5.29b)
$$
$$
X_{l,x}=\biggl[-{1\over 2}{g'\over g}+{1\over x}-{(n+1)\over
xg}\biggr] N_{l} +\biggl[{1\over 2}{g'\over g}-{1\over x}-{x\over
4}\biggl({g'\over g}\biggr)^2+ {x\over 2}{g''\over g}
$$
$$
+{(n+1)\over xg}-\sigma_{l}^2 {x\over g^2}\biggr] L_{l}+
\biggl[{1\over 2}{g'\over g}-{1\over x}- {x\over 4}\biggl({g'\over
g}\biggr)^2+{x\over 2}{g''\over g}- \sigma_{l}^2{x\over
g^2}\biggr] X_{l}.
                                                               \eqno(5.29c)
$$
This system can be transformed to the  equivalent form
$$
xg^2N_{l,x}=(n+1)gN_{l}+g\biggl({x\over 2}g'-(n+1)\biggr) L_{l}
$$
$$
+ x^2\biggl({1\over 4}(g')^2-{1\over 2}gg''+\sigma_{l}^2\biggr)
\tilde X_{l}
                                                                   \eqno(5.30a)
$$
$$
xgL_{l,x}+\biggl({x\over
2}g'+g\biggr)L_{l}=xgN_{l,x}+\biggl({x\over 2}g'-g\biggr)N_{l}
                                                                  \eqno(5.30b)
$$
$$
xg\tilde X_{l,x}=-gL_{l}+\biggl({x\over 2}g'-g\biggr)\tilde X_{l}
                                                                  \eqno(5.30c)
$$
where
$$
\tilde{X_l}=X_l+L_l,
                                                                \eqno(5.31)
$$
which can be compared with the analogous Chandrasekhar system
(\cite{chandra}, Ch.4, eqs. (46-47), (50)). Our equations
(5.30b)-(5.30c) coincide  with Chandrasekhar equations (46)-(47),
while our equation (5.30a) coincides with the Chandrasekhar
equation (50) if and only if $\rho^{\prime}=0$ which is equivalent
to $({x^2} g^{\prime\prime}/2-g+1)=0$.

The basic system (5.30) can be directly applied to study extreme black hole case. In
the next subsection we investigate first the case of a simple horizon to make clear
peculiarity of the case of the double horizon.

\subsection{Extreme black hole case}

{\bf Behavior near the simple horizon}

\vskip0.1in

In the neighborhood of a simple horizon
 $x_{+}$ we have $g(x)=g'(x_{+})(x-x_{+})+{1\over
 2}g''(x_{+})(x-x_{+})^2+...$. To study behavior in the limit $x \rightarrow x_{+}+0$
we introduce the variable $z=x-x_{+}$. In a small neighborhood of
$z=0$ limiting system for (5.30) reads
$$
x_{+}(g'(x_{+}))^2 z^2 N_{l,z}=(n+1)g'(x_{+})z N_{l}
$$
$$
+g'(x_{+})z\biggl[{x_{+}\over 2}g'(x_{+})-(n+1)\biggr]L_{l}+
x_{+}^2 \biggl({1\over
4}(g'(x_{+}))^2+\sigma_{l}^2\biggr)\tilde{X_l}
                                                                       \eqno(5.32)
$$
$$
z(N_{l}-L_{l})_{,z}+{1\over 2}(N_{l}-L_{l})=0
                                                                  \eqno(5.33)
$$
$$
x_{+}z \tilde X_{l,z}=-z L_{l}+{1\over 2}x_{+} \tilde X_{l}
                                                                      \eqno(5.34)
$$

 One immediately sees from (5.33) that the restricted near $z=0$
solutions should satisfy  $N_{l}(z)=L_{l}(z)$. Then (5.32) and
(5.34) form the system of two first-order equations for functions
$N_{l},L_{l}$
$$
(g'(x_{+}))^2 z^2 N_{l,z}={1\over 2}(g'(x_{+})^2 z N_{l}+ x_{+}
\biggl({1\over 4}(g'(x_{+}))^2+\sigma_{l}^2\biggr)\tilde X_{l},
                                                                       \eqno(5.35)
$$
$$
x_{+} z \tilde X_{l,z}=-zN_{l}+{1\over 2}x_{+} \tilde X_{l}
                                                                   \eqno(5.36)
$$
This system reduces to one second-order equation for  $\tilde
X_{l}$
$$
z^2\tilde X_{l,zz}-z\tilde X_{l,z}+\biggl(1+{\sigma_{l}^2\over
(g'(x_{+}))^2}\biggr)\tilde X_{l}=0
                                                                   \eqno (5.37)
$$
This is the Euler equation, and solutions of (5.32)-(5.34)
restricted near zero, have the form
$$
\tilde X_{l}(z)= \biggl[B_{1l}\cos\biggl({\sigma_{l}\over
g'(x_{+})}\ln z\biggr)+ B_{2l}\sin\biggl({ \sigma_{l}\over
g'(x_{+})}\ln z\biggr)\biggr]z,
                                                               \eqno (5.38)
$$
$$
 N_{l}=L_{l}=-x_{+}\biggr[\biggl({1\over
2}B_{1l}+ {\sigma_{l}\over g'(x_{+})}B_{2l}\biggr)\cos \biggl({
\sigma_{l}\over g'(x_{+})}\ln z\biggr)
$$
$$
+\biggl({1\over 2}B_{2l}-{ \sigma_{l}\over g'(x_{+})}B_{1l}\biggr)
\sin\biggl({\sigma_{l}\over g'(x_{+})}\ln z\biggr)\biggr],
                                                               \eqno(5.39)
$$
where $B_{1l},B_{2l}$ are arbitrary constants. As a result in the
small neighborhood of a simple horizon restricted solutions exist
for all real values of $\sigma_{l}$.

\vskip0.1in

{\bf Behavior near the double horizon}

\vskip0.1in

The double horizon $x_{\pm}$ corresponds to the case $\alpha=\alpha_{cr}$ in (5.28).
For the case of the density profile (1.4)
$$
\alpha_{cr} \simeq{1.456}
                                             \eqno(5.40)
$$

In the small neighborhood of the point
 $x=x_{\pm}$, the metric function is written as
$$
g(x)=\gamma(x-x_{\pm})^2+...,\gamma={1\over 2}g''(x_{\pm})
$$
In the  variable $z=x-x_{\pm}$,  in the small neighborhood of
$z=0$ the limiting system for (5.30) reads
$$
x_{\pm} \gamma^2 z^4 N_{l,z}=(n+1)\gamma
z^2(N_{l}-L_{l})+x_{\pm}^2 \sigma_{l}^2\tilde X_{l},
                                                          \eqno (5.41)
$$
$$
z(N_{l}-L_{l})_{,z}+(N_{l}-L_{l})=0,
                                                            \eqno (5.42)
$$
$$
x_{\pm} z \tilde X_{l,z}=-zL_{l}+x_{\pm}\tilde X_{l}
                                                            \eqno(5.43)
$$
As follows from (5.42), for a restricted solution it should be
$N_{l}=L_{l}$. Then equations (5.41) and (5.43) form a system of
two first-order equations for $N_{l},L_{l}$:
$$
z^4 N_{l,z}=x_{\pm} {\sigma_{l}^2 \over \gamma^2} \tilde X_{l},
                                                           \eqno (5.44)
$$
$$
z \tilde X_{l,z}=-{z\over x_{\pm}} N_{l}+ \tilde X_{l}
                                                          \eqno (5.45)
$$
This system reduces to one second-order equation for $N_{l}$
$$
z^4 N_{l,zz}+3z^3 N_{l,z}+ {\sigma_{l}^2\over \gamma^2} N_{l}=0
                                                         \eqno (5.46)
$$
which differs essentially from the analogous equation (5.37) for a
simple horizon case. General solution to (5.46) is \cite{kamke}
$$
N_{l}(z)={1\over z} \biggl[ C_{1l}J_{1}\biggl({ \sigma_{l}\over
\gamma}{1\over z}\biggr)+ C_{2l}Y_{1}\biggl({ \sigma_{l}\over
\gamma}{1\over z}\biggr)\biggr],
                                                        \eqno (5.47)
$$
 where $C_{1l},C_{2l}$
are arbitrary constants, $J_1, Y_1$ are Bessel functions.

With taking into account asymptotic behavior of Bessel functions
for big values of argument, we find the behavior of function
$N_{l}(z)$ for $z\rightarrow 0$

$$
N_{l}(z)={1\over z^{1\over 2}}\biggl[ C_{1l}\cos\biggl({
\sigma_{l}\over \gamma}{1\over z}- {3\pi\over 4}\biggr)+
C_{2l}\sin\biggl({\sigma_{l}\over \gamma}{1\over z}- {3\pi\over
4}\biggr)
                                                          \eqno (5.48)
$$

We see that solutions to (5.46) are unbounded as  $z\rightarrow 0$
 for all real values of the parameter $\sigma_l$.

From (5.44) we get
$$
 X_{l}(z)=-{\gamma z\over (x_{\pm})  \sigma_{l}}
\biggl[C_{1l}J_{0}\biggl({ \sigma_{l}\over \gamma}{1\over
z}\biggr) + C_{2l}Y_{0}\biggl({ \sigma_{l}\over \gamma}{1\over
z}\biggr)\biggr]
                                                              \eqno(5.49)
$$
which gives in the limit $z\rightarrow 0$
$$
X_{l}(z)= -{1\over (x_{\pm})} \biggl({2\over \pi}\biggr)^{1\over
2} \biggl({\gamma z \over \sigma_{l}}\biggr)^{3\over 2}
\biggl[C_{1l}\cos\biggl({ \sigma_{l}\over \gamma}{1\over z}-
{\pi\over 4}\biggr)
$$
$$
+ C_{2l}\sin\biggl({ \sigma_{l}\over \gamma}{1\over z} - {\pi\over
4}\biggr)\biggr]
                                                           \eqno (5.50)
$$

Analysis of our basic system (5.30) in small neighborhood of double horizon
$x=x_{\pm}$ shows that for all real values of the parameter $\sigma_l$, there exist
unbounded solutions as $x\rightarrow x_{\pm}$. Therefore the method of linear
perturbations, as well as the assumption (5.8), are not suitable in this case, but the
behavior of perturbations suggest instability of the extreme configuration. It should
be investigated separately, and we are currently working on this \cite{us2004}.

\subsection{The reduction of the system to a one-dimensional
wave equation}

Now our goal is to reduce the system (5.30) to a single second-order equation. We
introduce the new functions $z_{1l}, z_{2l}, z_{3l}$ using the linear transformations
$$
N_{l}=\biggl[{1\over x}z_{1l}+\biggl({g'\over 2}-{x\over 4 g}
(g')^2-\sigma_{l}^2 {x\over g}\biggr)z_{2l}+z_{3l}\biggr]g^{1\over
2},
                                                                    \eqno(5.51a)
$$
$$
 L_{l}=\biggl[\biggl(-{x\over 2}g''
 +{g'\over 2}\biggr)z_{2l}+z_{3l}\biggr]g^{1\over 2},
                                                                     \eqno(5.51b)
$$
 $$
 X_{l}=\biggl[\biggl({(n+1)\over x}-{g\over x}+ {x\over
2}g''\biggr)z_{2l}-z_{3l}\biggr]g^{1\over 2}
                                                                    \eqno(5.51c)
$$
The inverse transformation to
 (5.51) reads
$$
 z_{1l}(x)=\frac{x}{\sqrt{g} b}\biggl[b(x)N_{l}(x)+\biggl((b(x)-n-1)\frac{x}{2}\frac{g'}{g}-xb^{\prime}
$$
$$
-b(x) +\sigma_{l}^2 \frac{x^2}{g(x)}\biggr)L_{l}+
\biggl((b-n-1)\frac{x^2}{2}\frac{g'}{g}-xb_{,x}+\sigma_{l}^2
\frac{x^2}{g(x)}\biggr)X_{l}\biggr],
                                                                    \eqno (5.52a)
$$
$$
 z_{2l}(x)=[L_{l}+X_{l}]\frac{x}{\sqrt{g} b(x)},
                                                                  \eqno(5.52b)
$$
$$
 z_{3l}(x)=\biggl[\biggl({1\over x}b(x)+b_{,x}\biggr)L_{l}+b_{,x}X_{l}\biggr]
\frac{x}{\sqrt{g} b(x)}
                                                                   \eqno (5.52c)
$$
where
$$b(x)=n+1+{x\over 2}g'(x)-g(x)=n+\frac{3\alpha}{2x}\biggl(M(x)-x^3\rho\biggr)
                                                                    \eqno(5.53)
$$
The sum of (5.51b) and (5.51c) gives
$$
{X}_{l}+L_l=\biggl[\frac{n+1}{x}+\frac{1}{2} g' - \frac{1}{x}
g\biggr] g^{1/2} z_{2l}
                                                                   \eqno(5.54)
$$
 As a result we get the following system
$$
z_{1l,x}=\biggl({2\over x}-{g'\over g}\biggr)z_{1l}-\biggl({1\over
2}x^2g'''+xg''-g'\biggr)z_{2l}
$$
$$
+\biggl[2+{x^2\over b(x)}\biggl({g''\over 2}- \frac{(g')^2}{4g}
-\sigma_{l}^2 \frac{1}{g}\biggr)\biggr]z_{3l},
                                                                  \eqno(5.55a)
$$
$$
z_{2l,x}=-{1\over b(x)}z_{3l},
                                                                  \eqno(5.55b)
$$
$$
 z_{3l,x}={b(x)g^{-1}\over x^2}z_{1l}-
\biggl[{2\over x}+{(xg''-g')\over 2b(x)}\biggr]z_{3l}$$
$$ +{1\over
x}\biggl({{x^2} \over 2}g'''+xg''-g'\biggr)z_{2l},
                                                                   \eqno(5.55c)
$$

It is easily to prove that
$${1\over 2}x^2g'''+xg''-g'=
-{3\alpha\over 2}(x^3\rho^{\prime})^{\prime} =3\alpha x^2
p_{\perp}^{\prime}
                                                                  \eqno(5.56)
$$
so that in the case when the density profile satisfies the
condition
$$
(x^3\rho^{\prime})^{\prime}=0,
                                                                \eqno(5.57a)
$$
the system (5.55) splits on the system of two equations (5.55a),
(5.55c) for  $z_{1l}, z_{3l}$, and the equation (5.55b).

Condition (5.57a) is in turn equivalent to
$$
p_{\perp}^{\prime}=0
                                                                  \eqno(5.57b)
$$

In the particular case $(x^3\rho^{\prime})=const=0$ this is the necessary and
sufficient condition for coinciding of our system (5.30) with the Chandrasekhar system
(\cite{chandra}, eqs.(46)-(47), (50) Ch.4).

Differentiating (5.55c), we come to the system which includes one
first-order equation, (5.55b), and one second-order equation
$$
z_{3l,xx}+2\biggl({g'\over g}+{1\over x}\biggr)z_{3l,x}+
{q}_{l}(x) z_{3l}= r_{l}(x) z_{2l},
                                                                \eqno(5.58)
$$
where
$$
q_{l}(x)= \sigma_{l}^2{1\over g^2} -{2(n+1)\over x^2g}-{1\over
2}{g''\over g}+{1\over 4} \biggl({g'\over g}\biggr)^2
+{3\over x}{g'\over g}
$$
$$
-{(xg''-g')\over b(x)}\biggl[ {(xg''-g')\over 2b(x)}-{g'\over
g}+{1\over x}\biggr] +{xg'''\over 2b(x)}
 + {3\alpha x\over b(x)}p_{\perp}^{\prime}
                                                          \eqno(5.59)
$$
$$
r_{l}(x)= -3\alpha p_{\perp}^{\prime} \biggl[{(n+1)\over
g}-{3x\over 2}{g'\over g}
$$
$$
+{x\over 2b(x)} (xg''-g')\biggr] +{3\alpha\over x}(x^2
p_{\perp}^{\prime})^{\prime}.
                                                                \eqno(5.60)
$$

It is easily to see that in the case when the condition (5.57) is
satisfied,  two equations (5.55b) and (5.58) split. Introducing
the new function $\omega_{3l}(x)$ by
$$
z_{3l}(x)={1\over xg}\omega_{3l}(x)
                                                                   \eqno(5.61)
$$
we reduce the equation (5.58) to the form which does not contain the first derivative:
$$
\omega_{3l,xx}+\biggl[\sigma_{l}^2 {1\over
g^2}-V_{1l}(x)\biggr]\omega_{3l}= xg r_{l}z_{2l},
                                                                  \eqno(5.62)
$$
where the potential $V_{1l}(x)$ is given by
$$
V_{1l}(x)={l(l+1)\over x^2}{1\over g}+ {3\over 2}{g''\over g}-
{1\over x}{g'\over g}+ {(xg''-g')\over b(x)}
\biggl[{(xg''-g')\over 2b(x)}
$$
$$
-{g'\over g}+{1\over x}\biggr]-{1\over 4}\biggl({g'\over
g}\biggr)^2 - {xg'''\over 2b(x)}-{1\over xb(x)}\biggl[{1\over
2}x^2g''' +xg''-g'\biggr]
                                                                      \eqno(5.63)
$$
 With taking into account (5.55b) and (5.61), equation (5.62)
 can be rewritten as integro-differential equation of the form
$$
\omega_{3l,xx}+\biggl[g^{-2}(x)\sigma_{l}^2 -
V_{1l}(x)\biggr]\omega_{3l}(x)
$$
$$
=-xg(x) r_{l}(x)\int{{\omega_{3l}(x) \over xg(x)b(x)}dx}
                                                                      \eqno(5.64)
$$
In this form ($g^{-2}(x)$ scales the spectral parameter
$\sigma^2$) the equation (5.64) corresponds to the generalized
spectral problem with the non-local potential
$$
-\omega_{3l,xx}+V_{1l}(x)\omega_{3l}(x) -T_{l}\omega_{3l}(x)
=\sigma_{l}^2 {1\over g^2}\omega_{3l}(x),
                                                                       \eqno(5.65)
$$
where
$$
T_{l}u(x)=xg(x) r_{l}(x) \int\limits_{d}^{x}{{u(z)dz\over z b(z)
g(z)}}
                                                                          \eqno(5.66)
$$
 is the integral Vol'terra operator. The lower limit is $d=x_{+}$
 for a black hole case.

The condition (5.57) (which leads to $r_l=0$) is necessary and sufficient condition to
reduce (5.64) to the  Schr\"odinger equation with the local potential.

Introducing "the tortoise coordinate" $x_{*}(x)=\int{dx/g(x)}$ and
the function $w(x_{*})$ by
$$
w_{3l}(x_{*})=x \sqrt{g(x)}z_{3l}(x_{*})
                                                                   \eqno(5.67)
$$
we reduce the system (5.58), (5.55b) to the form
$$
w_{3l,x_{*}x_{*}}+[\sigma_{l}^2 -W_{l}(x)] w_{3l}(x_{*})=
xg^{{5\over 2}}(x) r_{l}(x) z_{2l}(x_{*})
                                                                   \eqno(5.68)
$$
$$
z_{2l,x_{*}}=-{g^{1/2}(x)\over xb(x)} w_{3l}(x_{*}),
                                                                   \eqno(5.69)
$$
where
$$
W_{l}(x)=g\biggl[{l(l+1)\over x^2}+g''-{1\over x}g'
+{g(xg''-g')\over b}\biggl({{xg''-g'}\over 2b}
$$
$$
-{g'\over g} +{1\over x}\biggr) -{xgg'''\over 2b}-{g\over
xb}\biggl({1\over 2}x^2g'''+ xg''-g'\biggr)\biggr]
                                                                    \eqno(5.70)
$$
In the limit $x\rightarrow x_{+}$ the integral term in (5.66) tends to zero, on
essence due to $z_{2l}\rightarrow 0$. Indeed, when $x\rightarrow x_{+}$, we get
$z_{2l} \sim {\sqrt{x-x_{+}}}$ with using (5.52b) and taking into account asymptotic
behavior of $(L_l+X_l)\sim{(x-x_{+})}$ which follows from (5.38).

The potential (5.70) vanishes as $x_{*}\rightarrow +\infty$ as $x^{-2}$, while for
$x_{*}\rightarrow -\infty$, it vanishes exponentially. Therefore solutions to (5.68)
have asymptotic $e^{\pm i\sigma_l x_{*}}$ as $x_{*}\rightarrow \pm \infty$, so that we
have to look for solutions satisfying boundary conditions
$$
w_{l} \rightarrow e^{i\sigma_l x_{*}} + R_l^{(w)} e^{-i\sigma_l
x_{*}}  \quad{as~~~} x_{*}\rightarrow \infty
$$
$$
w_{l} \rightarrow T_l^{(w)}e^{i\sigma_l x_{*}} \quad{as~~~}
x_{*}\rightarrow -\infty
                                                         \eqno(5.71)
$$
In the particular case of validity of (5.57), $ r_{l}(x)=0$,
$p_{\perp}^{\prime}=0$, the system (5.68)-(5.69) splits and we get
the Schr\"odinger equation
$$
-w_{3l,x_{*}x_{*}}+W_{0l}(x) w_{3l}(x_{*})= {\sigma}_{l}^2
w_{3l}(x_{*})
$$
with the potential
$$
W_{0l}(x)=g\biggl[{l(l+1)\over x^2}+g''-{1\over x}g'
$$
$$
+{g(xg''-g')\over b(x)}\biggl({{xg''-g'}\over 2b(x)}-{g'\over g}
+{2\over x}\biggr)\biggr],
$$
which for the Schwarzschild geometry coincides with the potential
in the  Zerilli equation (\cite{chandra}, Ch.4, eq.(63)).

\vskip0.1in

We have reduced the basic system of three first-order linear equations (5.29) to a
single second-order equation for the particular combination of these function,
$w_{3l}(x_*)$. In our case this is the Schr\"odinger equation (5.72) with non-local
potential. Its non-local part vanishes when the condition (5.57) is satisfied. As in
Schwarzschild case, this empirically found reducibility (resulted from a sequence of
mysterious cancellations), follows in fact from the existence of some particular
solution to the system which we have to reduce \cite{chandra}.

In our case the condition (5.57) is the necessary and sufficient condition for
vanishing non-local part in (5.72) and thus for reducing  our system (5.29) to the
standard Schr\"odinger equation. Actually, the condition (5.57) is also the necessary
and sufficient condition for existence of  particular solution which guarantees such a
reduction (it is shown in detail in our paper \cite{us2004} devoted to the case
satisfying (5.57)). If the condition (5.57) is satisfied, then the particular solution
reads
$$
N_l^p=\sqrt{g}\biggl[\frac{g"}{2g}-\biggl(\frac{g'}{2g}\biggr)^2
-\sigma_l^2\frac{x}{g}-b'(x)\biggr]
$$
$$
L_l^p=-\sqrt{g}b'(x);~~X_l^p=\sqrt{g}\biggl(\frac{b(x)}{x}+b'(x)\biggr)
$$

The linear transformation (5.51) is needed to reduce the system of 3 linear equations
to one equation, if the particular solution to (5.29) exists. Remarkable luck in our
case is that this transformation works not only in the case when (5.57) is satisfied
(i.e. obtained second order equation is the standard Schr\"odinger equation with the
local potential), but also in general case. This fact allowed us to reduce our system
(5.29) to the Schr\"odinger equation (5.72) with the non-local potential.

We see that the problem of stability of the spherically symmetric
solution (2.7) to polar perturbations, reduces to investigation of
the spectral problem (5.68)-(5.69) with the potential of the form
(5.70) and with the boundary conditions (5.71).

Equation (5.68), with taking into account (5.69), can be written
in the form corresponding to the spectral problem with a non-local
potential
$$
-w_{3l,x_{*}x_{*}}+ W_{l}(x)w_{3l}(x_{*})- \tilde
T_{l}w_{3l}(x_{*}) = \sigma_{l}^2 w_{3l}(x_{*}),
                                                                        \eqno(5.72)
$$
where
$$
\tilde T_{l}u(x_*)= x(x_*) g^{{5\over 2}} (x(x_*)) r_{l} (x(x_*))
\int\limits_{d_*}^{x_*} {{g^{1\over 2} (x(z_*)) u(z_*)dz_*\over
{x(z_*)b(x(z_*)) }}},
                                                                      \eqno(5.73)
$$
is the integral Vol'terra operator.

Ultimately the case reduces to investigation of the eigenvalue
problem with integro-differential operator (5.72), which is the
Schr\"odinger equation with the non-local potential.

The spectrum of eigenvalues contain all the values of the
parameter ${\sigma}_{l}^2$, at which solutions exist which satisfy
the imposed boundary conditions.

If such solutions exist only for real values of the time parameter
${\sigma}_{l}$ and if, in addition, they form the complete basic
set of functions, then any smooth initial perturbation on the
finite interval of variable $x_{*}$ (with compact measure) can be
expanded on this basic set, and since dependence of each
particular mode  on time is given by $\exp(i{\sigma}_{l}t)$, it
testifies for stability of geometry.

Indeed, a considered static configuration is stable if there are
no integrable modes with negative $\sigma_l^2$. Appearance of
negative eigenvalues $\sigma_l^2$ would lead to existence of
exponentially growing modes of perturbations.

In the next subsection we study in detail the integro-differential
operator governing polar perturbations.

\subsection{Schr\"odinger equation with non-local potential}

A system governed by Schr\"odinger equation with a non-local
potential, obeys the following theorems:

i) If in the standard one-dimensional Schr\"odinger equation the potential is
nonnegative, then negative eigenvalues are absent (see, e.g., \cite{ber}).

ii) The Weyl theorem \cite{weyl} for self-conjugate operators:

The essential spectrum conserves under relatively compact perturbations \cite{kato}.

The essential spectrum is defined as follows:
 If we remove from the spectrum of self-conjugate
operator all isolated points which are eigenvalues of finite multiplicity, then the
remaining of the spectrum is the essential spectrum.

Essential spectrum of non-perturbed (local) potential is continuous and represented by
positive semi-axis $[0, \infty)$, isolated points are absent in case when negative
values are excluded by non-negativity of the potential.

It follows that the essential spectrum of the problem with the non-local potential
(5.72) is the same as the essential spectrum of the non-perturbed (local) potential.
Essential spectrum of local potential is this total spectrum, since isolated points
are absent, i.e., essential spectrum of the perturbed problem coincides with the total
spectrum of the non-perturbed problem.

Non-local part of a potential represents the perturbation of the local potential. To
not spoil an essential spectrum, this perturbation should be relatively compact.

So, our  task  now is to prove that non-local part represents a compact perturbation
and to deduce criterion of non-negativity of a local potential.

 In our case non-local part (perturbation of a local potential)
 is given by the integral Vol'terra operator (5.73). Such an operator
 is totally continuous, if it has smooth square integrable kernel.

Square integrability requires
$$
\int_{-\infty}^{\infty}\int_{-\infty}^{\infty}{K^2(x_*, z_*) dx_*
dz_*} < \infty
                                        \eqno(5.74)
$$
The kernel of our Vol'terra operator (5.73) is
$$
K(x_*, z_*)= x(x_*) g^{{5\over 2}} (x(x_*))  r_{l} (x(x_*))
 {{g^{1\over 2} (x(z_*)) \over {x(z_*)b (x(z_*))
}}}.
                                                                    \eqno(5.75)
$$
Its smoothness is evident. The sufficient condition for square integrability of the
kernel $K(x,y)$ is the condition on behavior of $K^2$ at infinity:
$$
K^2(x,y) < \frac{1}{x^{1+\delta_1}}\frac{1}{y^{1+\delta_2}}
$$
where $\delta_1, \delta_2$ are arbitrarily small.

For $K(x_{*}, z_{*})$ given by (5.75), for $z_* \rightarrow
{-\infty}$,  $K^2$ vanishes as $g(x(z_*))$. When $x_* \rightarrow
{-\infty}$, then $K^2$ vanishes as $g^3(x(x_*))$. The metric
function $g(x(x_*))$ near horizon behaves as
$g(x_*)\simeq{g'(x_*)e^{g'(x_*) x_*}}$, $g'$ is positive, so that
metric as a function of $x_{*}$  vanishes exponentially at
approaching the horizon.

When $z_* \rightarrow \infty$, then $K^2$ vanishes as $x^{-2}(z_*)$. From definition
$z_*$ we see the main contribution at infinity is $z_*\sim{x}$. When $x_* \rightarrow
\infty$, $K^2$ vanishes as $x^2 (p_{\perp}')^2$. The tangential pressure $p_{\perp}$
for de Sitter-Schwarzschild geometry vanishes at infinity quicker than $x^{-3}$,
because it is related with density by the equation of state
$p_{\perp}=-\rho-x{\rho}'/2$, and density vanishes quicker than $x^{-3}$ to guarantee
the finiteness of a mass. Hence in this limit $K^2$ vanishes quicker than $x^{-6}$.

So, for a BH case the kernel is square integrable.

As a result the totally continuous operator (5.73) gives a relatively compact
perturbation to the  local potential in the integro-differential equation (5.72).

\vskip0.1in

{\bf Criterion of non-negativity of  local potential}

Introducing the function $p(x)=xg''(x)-g'(x)$, we write the
potential (5.72) in the form
$$
W_{l}(x)=g\biggl[{1\over2}g\biggl({p\over b}\biggr)^2+{1\over
2b(x)}(g')^2+ {2(n+1)\over x^2} -{1\over bx} I_{l}(x)\biggr],
                                                                       \eqno(5.76)
$$
where
$$
I_{l}(x)=x^2\biggl({1\over 2}g'g''+gg'''\biggr)-(n+1-g)p(x)
                                                                    \eqno (5.77)
$$
In (5.76) we should investigate the term $I_{l}$. The rest is positive, since
$b(x)\geq n$ in a BH case.

Expressing $g(x)$,  its derivatives, $p(x)$ and $b(x)$ in terms of
mass function $M(x)$, density $\rho(x)$ and its derivatives, we
transform $I_{l}(x)$ to the form
$$
I_{l}(x)=\alpha\biggl[-4\alpha{M^2\over x^3}+ {9\over 2}\alpha
x^4\rho\rho^{\prime}+ 3(n-1)x^2\rho^{\prime}
$$
$$
+ {9\over 2}\alpha x\rho^{\prime}M-3nx\rho
-3gx(x^2\rho^{\prime\prime}+2\rho)+ {3(n+2)\over x^2}M\biggr]
                                                                   \eqno(5.77a)
$$
For a BH case $g(x)>0$ while the weak energy condition gives
$\rho^{\prime}<0$. Then the sufficient condition for $W_l\geq 0$
is the condition on the equation of state
$$
x^2\rho^{\prime\prime}(x)+2\rho(x)\geq 0,
                                                                  \eqno (5.78a)
$$
which constraints the growth of the
 derivative of  $p_{\perp}+\rho$
 $$
 x(p_{\perp}+\rho)^{\prime}\leq \rho + (p_{\perp}+\rho)
                                                                \eqno(5.78b)
 $$
This condition is actually satisfied also for the case without
horizons (then $g(x)>0$ for all $x$).

 When (5.78) is satisfied, then proof of non-negativity of (5.76) reduces
 to proof of non-negativity of the function
$$
\phi(x)={2(n+1)\over x^2}-{3\alpha(n+2)M\over x^3 b(x)}
                                                                 \eqno (5.79)
$$
It is bounded from below as follows
$$
\phi(x)={2\over x^2}\biggl[n+1-{3\alpha M(n+2)\over 2 x
b(x)}\biggr]
$$
$$
\geq{2\over x^2}\biggl[n+1-{3\over 2}{(n+2)\over n}{\alpha M\over x}\biggr] ={2\over
x^2}\biggl[n+1-{3\over 2}{(n+2)\over n}(1-g)\biggr]
$$
$$
> {2\over
x^2}\biggl[n-{1\over 2}-{3\over n}\biggr]\geq\frac{2}{x^2}(n-2)\geq 0.
                                                                \eqno (5.80)
$$

As a result, we find the sufficient condition (5.78) for
non-negativity of the potential (5.70) in all range of argument
for which $g(x)>0$.

{\it For  the density profile (1.3) this condition is satisfied.}

We can conclude that the essential spectrum of the
integro-differential operator (5.72) is the same as the essential
spectrum of its local potential. Now the
key point is to find the condition on a perturbation of a local
potential which guarantees the absence of isolated points in the
total spectrum (negative values of $\sigma_l^2$~) of the
integro-differential operator (5.72).

\vskip0.1in

{\bf Non-local contribution}

Multiplying (5.68) by $w_{3l}^{*}$\footnote{We denote the complex conjugate by $^*$
for convenience of comparison with the classical results presented in \cite{chandra}.}
and integrating by parts with taking into account asymptotic behavior of (5.60) at
infinity, we obtain the following relation
$$
\sigma_{l}^2 \int\limits_{-\infty}^{+\infty}{|w_{3l}(x_{*})|^2}dx_{*} +
w_{3l,x_*}w_{3l}^*|_{-\infty}^{\infty}=
$$
$$
\int\limits_{-\infty}^{+\infty}[ |w_{3l,x_{*}}|^2+W_{l}(x)|w_{3l}(x_{*})|^2+
\psi_{l,x} g(x) |z_{2l}(x_{*})|^2]dx_{*}
                                                                          \eqno (5.81)
$$
where
$$
\psi_{l}(x)=\frac{x^2}{2} g^2(x) b(x) r_{l}(x)
                                                      \eqno(5.82)
$$
The Wronskian $w_{3l,x_*}w_{3l}^*|_{-\infty}^{+\infty}$ of two independent solutions
$w_{3l}$ and $w_{3l}^*$ is constant (see \cite{chandra}, Par.27 Ch.4).

 The contribution to the spectrum from the non-local
part of the potential is given by

$$
N= \int\limits_{x_{+}}^{+\infty}\psi_{l,x}(x)|z_{2l}(x)|^2dx
                                                                             \eqno (5.83)
$$
 If the condition of non-negativity of a local potential $W_{l}(x)$
is satisfied, then the requirement
$$
N= \int\limits_{x_{+}}^{+\infty}\psi_{l,x}(x)|z_{2l}(x)|^2dx\geq 0
                                                                             \eqno (5.84)
$$
gives the sufficient condition for the absence of negative
eigenvalues $\sigma_l^2$ of the considered spectral problem.

Fortunately, non-local
 contribution given by (5.83), does not grow with
 the mode number $n$, since $|z_{2l}|^2$ is constrained from above
 by the function proportional to $n^{-2}$. This constraint is valid for any density
 profile and follows from (5.69) with taking into account that in a BH case $b(x)\geq n$.

\vskip0.1in

In the case when the metric function $g(x)$ satisfies the
condition (5.57), the sufficient condition (5.84) is trivially
satisfied ($N=0$), and negative eigenvalues do not appear in the
spectrum. As a result spherically symmetric metrics satisfying
(5.57) and (5.78) are stable to polar perturbations.

 \section{The case of density profile (1.3)}

In the case of the density profile $\rho(x)=e^{-x^3}$ the metric function in (5.28)
reads
$$
g(x)=e^{2\nu(x)}=1-{\alpha\over x}(1-e^{-x^3})
                                                            \eqno(6.1)
$$

Potential (4.14) governing the axial perturbations is depicted in Figs.5-6 for the
density profile (1.3) and two values of the characteristic parameter $\alpha$ (denoted
in figures as $a$).
\begin{figure}
\vspace{-8.0mm}
\begin{center}
\epsfig{file=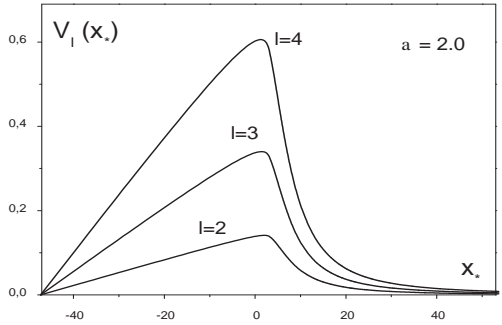,width=8.0cm,height=5.5cm}
\end{center}
\caption{Axial potential (4.14) for $m\simeq{2.8m_{cr}}$.}
\label{fig.5}
\end{figure}
\begin{figure}
\vspace{-8.0mm}
\begin{center}
\epsfig{file=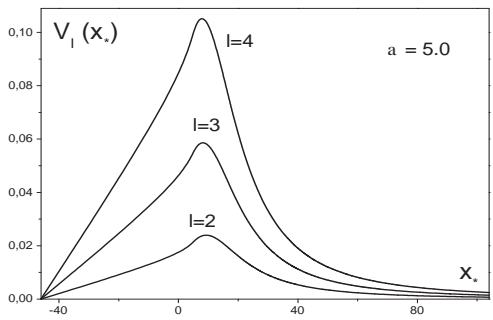,width=8.0cm,height=5.5cm}
\end{center}
\caption{Axial potential (4.14) for $m\simeq{11.2m_{cr}}$.}
\label{fig.6}
\end{figure}
Potentials are smooth, short-range and positive for all values of
the characteristic parameter $\alpha$. Therefore all types of
vacuum configurations with de Sitter center including a vacuum
nonsingular black hole, are stable to axial perturbations.

 The local  potential governing polar perturbations given
by (5.70)  is shown in figs.7-8 for the density profile (1.3) and two values of
parameter $\alpha$ (denoted in  figures as $a$).
\begin{figure}
\vspace{-8.0mm}
\begin{center}
\epsfig{file=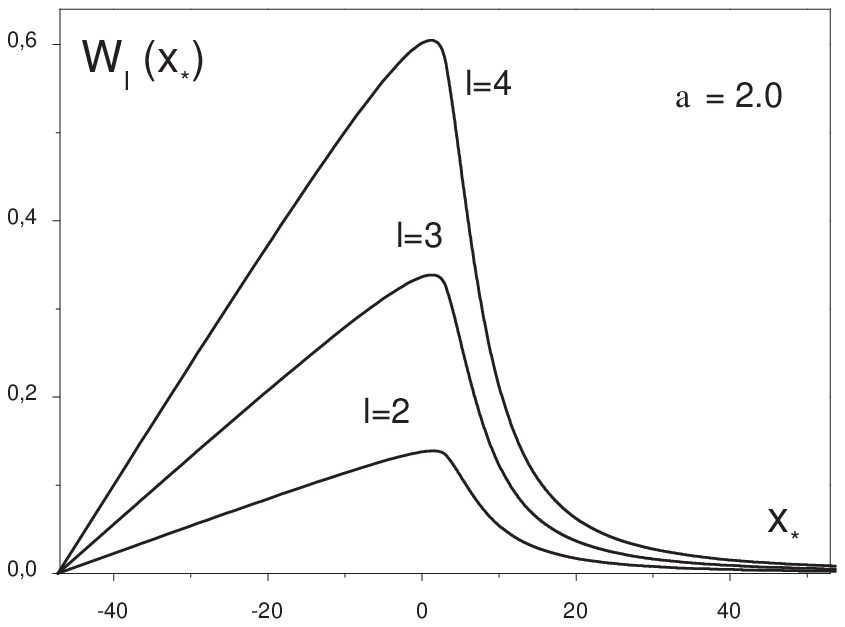,width=8.0cm,height=5.5cm}
\end{center}
\caption{Polar potential (5.70) for  $m\simeq{2.8 m_{crit}}$. } \label{fig.7}
\end{figure}
%
\begin{figure}
\vspace{-8.0mm}
\begin{center}
\epsfig{file=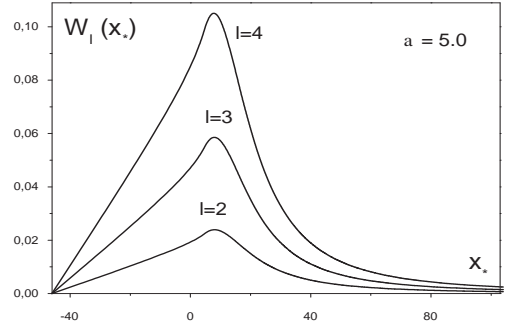,width=8.0cm,height=5.5cm}
\end{center}
\caption{Polar potential (5.70) for  $m\simeq{11.2 m_{crit}}$. } \label{fig.8}
\end{figure}

Both axial and polar potentials for bigger values of $\alpha$ become similar to those
for the Schwarzschild case \cite{chandra}.

Polar local potentials are smooth short-range potentials, so that
integrals of them are finite over all the region of variable $x$.

The potential (5.70) for the density profile (1.3) satisfies the
criterion of non-negativity (5.78), but the condition (5.57) is
not satisfied, so that appearance of negative eigenvalues
$\sigma_l^2$ is in principle possible.

 \vskip0.2in

The question of existence of isolated points with negative values $\sigma_l^2$ in the
total spectrum of the integro-differential operator(5.72) requires the complicated
numerical analysis which is in progress. Preliminary results suggest that non-local
contribution (5.83) does not lead to negative values $\sigma_l^2$ for the masses  $m >
m_{cr2}$.

This result looks natural. The second critical mass value $m_{cr2}$ is distinguished
for the unperturbed geometry. The value $m_{cr2}$ marks the point in the
temperature-mass diagram at which specific heat is broken and changes its sign, so
that a second order phase transition starts when in the course of Hawking evaporation
the mass approaches  the value $m_{cr2}$. For the density profile (1.3) it is given by
$$
m_{cr2}\simeq{0.38 m_{Pl}\sqrt{\rho_{Pl}/\rho_0}}
                                                                        \eqno(6.2)
$$
The extreme state of non-singular black hole ($m=m_{cr}$), can be unstable since some
perturbations modes grow unlimited at the double horizon for any density profile. If
the considered configuration would develop instability before achieving the extreme
state, the most appropriate range gets beyond $m_{cr2}$ where a phase transition
starts.

The critical value  $m_{cr2}$ corresponds to the maximum at the
temperature-mass curve (see fig.3). It is calculated from the
condition $dT/dm=0$. In the units normalized to de Sitter radius
$r_0$ (which is the characteristic scale related to de Sitter
vacuum trapped in the origin)
$$
y_{+}=\frac{r_{+}}{r_0}; ~~~ s=\frac{r_g}{r_0}
                                                   \eqno(6.3)
$$
 the temperature on a BH event horizon is
given by \cite{me96}
$$
T=\frac{1}{y_{+}}-\frac{3}{y_{+}}\biggl(1-\frac{y_{+}}{s}\biggr)
                                                                \eqno(6.4)
$$
The density profile and metric in these units read
$$
\rho(y)=e^{-y^3/s}; ~~ g(y)=1-{s\over y}\biggl(1-
e^{-y^3/s}\biggr)
                                                                       \eqno(6.5)
$$
From $dT/ds=0$ and $g(y_{+})=0$, we get the critical value $s_{2}$
and the value $y_{+}$ corresponding to $m=m_{cr2}$.
$$
s_2=\simeq 2.226; ~~~~y_{+}=\simeq{2.166}
                                                                    \eqno(6.6)
$$
For comparison, the critical values for the extreme case $m_{cr}$
of the double horizon $r_{\pm}$ are \cite{me96}
$$
s_{cr}\simeq{1.7576};~~~y_{\pm}\simeq{1.4957}
                                                          \eqno(6.7)
$$

\section{Discussion}

We present the conditions specifying two types of configurations with de Sitter
center, including black holes with and without changes of topology.

We found  that any configuration described by spherically
symmetric geometry with de Sitter center is stable to axial
perturbations.

The problem of stability to polar perturbation reduces to a
one-dimensional Schr\"odinger equation with a non-local potential
given by the Vol'terra integral operator with square integrable
smooth kernel representing a compact perturbation to the local
potential.

We derived the criterion of non-negativity of the local potential which defines the
essential spectrum of integro-differential operator governing polar perturbations in
general case.

We derived the criterion of vanishing of non-local part of the
potential which distinguishes the class of geometries for which
the problem of stability reduces to a standard one-dimensional
Schr\"odinger equation. This class of metrics has been studied in
\cite{us2004}.

For the case when perturbations are described by the Schr\"odinger equation with the
non-local potential, we found the sufficient condition for the absence of the negative
eigenvalues in the spectrum which guarantees the stability of investigated geometry.

 For an extreme black hole, the method of small perturbations
 is not applicable due to existence of unlimited perturbation modes
 at approaching the double horizon.

 Asymptotic behavior
 of the basic system near double horizon suggests instability of the extreme configuration.
 The behavior in this regime is very special, unrestricted solutions for perturbations
 exist for positive values of the spectral parameter $\sigma_l^2$. The limiting equation
 for perturbations near double horizon is essentially different from that for one-horizon
 case in which unrestricted solutions do not appear for positive values $\sigma_l^2$
 and which cannot be smoothly continued to two-horizon case. The question arises what
 is the place of the metric with $m=m_{cr}$, at the set of metrics whose stability
 we investigate as one-parametric set of solutions to Einstein unperturbed
 equations with a given density profile.

The critical value of the mass parameter $m_{cr}$ is calculated from two
transcendental equations: $g(r_{\pm})=0;~g'(r_{\pm})=0$. This is the unique point for
considered metric function $g(r)$ because it is the minimum of $g(r)$ and $g(r)$ has
the only one minimum \cite{me2002}. Therefore the transcendental system for $(r_{\pm},
m_{cr})$ has the unique solution for each particular one-parametric set with a given
density profile.

The metric with  double horizon represents  an isolated singular point at the set of
metrics $g(r)$ for each given density profile. It resembles a fixed point attractor
behavior which is currently the key point of our efforts \cite{us2004}.

\section* {Acknowledgement}

This work was supported by the Polish Committee for Scientific Research through the
grant 5P03D.007.20, at the final stage through the grant 1P03D.023.27 and by
St.Petersburg Scientific Center through the grant 9S22.

\end{document}